\documentclass[showpacs,twocolumn,pra,superscriptaddress,nofootinbib]{revtex4-1}
\pdfoutput=1

\usepackage[dvipsnames]{xcolor}

\usepackage{booktabs}
\usepackage{enumitem}
\usepackage{bm}
\usepackage{amsmath}
\usepackage{amssymb}
\usepackage{graphicx}
\usepackage{cleveref}
\usepackage{slashed}
\usepackage{amsfonts}
\usepackage{amsbsy}
\usepackage{color}
\usepackage{array}
\usepackage{dcolumn}
\usepackage{tensor}
\usepackage{braket}
\usepackage{mathrsfs}
\usepackage{tikz}
\usepackage{verbatim}

\newcommand{\sfP}{\mathsf{P}}

\newcommand{\sfW}{\mathsf{W}}

\newcommand{\mcE}{\mathcal{E}}

\newcommand{\mcN}{\mathcal{N}}
\newcommand{\mcO}{\mathcal{O}}

\newcommand{\ud}{{\textrm{d}}}
\newcommand{\lcm}{{\scriptscriptstyle -}} 
\newcommand{\lcp}{{\scriptscriptstyle +}}

\newcommand{\lcperp}{{\scriptscriptstyle \perp}}

\newcommand{\Ai}{\text{Ai}}

\begin{document}

\title{
    All-optical nonlinear Breit-Wheeler pair production with $\gamma$-flash photons
}

\author{Alexander J.~MacLeod}
\email{alexander.macleod@eli-beams.eu}
\affiliation{ELI Beamlines Centre, Institute of Physics, Czech Academy of Sciences, Za Radnic\`{i} 835, 25241 Doln\`{i} B\v{r}e\v{z}any, Czech Republic}

\author{Prokopis Hadjisolomou}
\affiliation{ELI Beamlines Centre, Institute of Physics, Czech Academy of Sciences, Za Radnic\`{i} 835, 25241 Doln\`{i} B\v{r}e\v{z}any, Czech Republic}

\author{Tae Moon Jeong}
\affiliation{ELI Beamlines Centre, Institute of Physics, Czech Academy of Sciences, Za Radnic\`{i} 835, 25241 Doln\`{i} B\v{r}e\v{z}any, Czech Republic}

\author{Sergei V. Bulanov}
\affiliation{ELI Beamlines Centre, Institute of Physics, Czech Academy of Sciences, Za Radnic\`{i} 835, 25241 Doln\`{i} B\v{r}e\v{z}any, Czech Republic}
\affiliation{National Institutes for Quantum and Radiological Science and Technology (QST), Kansai Photon Science Institute, 8-1-7 Umemidai, Kizugawa, Kyoto 619-0215, Japan}

\begin{abstract}
    High-power laser facilities give experimental access to fundamental strong-field quantum electrodynamics processes.
    A key effect to be explored is the nonlinear Breit-Wheeler process: the conversion of high-energy photons into electron-positron pairs through the interaction with a strong electromagnetic field.
    A major challenge to observing nonlinear Breit-Wheeler pair production experimentally is first having a suitable source of high-energy photons.
    In this paper we outline a simple all-optical setup which efficiently generates photons through the so-called $\gamma$-flash mechanism by irradiating a solid target with a high-power laser.
    We consider the collision of these photons with a secondary laser, and systematically discuss the prospects for exploring the nonlinear Breit-Wheeler process at current and next-generation high-power laser facilities. 
\end{abstract}

\maketitle

\section{Introduction \label{sec:Intro}}

Modern advances in laser technology have brought us into the multi-PW laser power regime, with a large number of high-power laser facilities~\cite{Danson.HPLSE.2019} either operational or in development, e.g.~\cite{Kiriyama.OL.2018,Sung.OL.2017,Yoon.Optica.2021,Nees.CLEO.2021,Hernandez-Gomez.JoP.2010,Papadopoulos.HPLSE.2016,Weber.MRE.2017,Gales.RPP.2018,Bromage.HPLSE.2021,Gan.2021,Zuegel.CLEO.2014,Mukhin.QE.2021,Shen.PPCF.2018}.
High-power lasers generate intense electromagnetic fields, allowing access to the nonlinear regime of quantum electrodynamics (QED), where the interaction between particles and laser fields cannot be described by the usual methods of vacuum perturbation theory.
Instead, the electromagnetic field must be taken into account non-perturbatively through a framework typically referred to as strong-field QED~\cite{Ritus.1985,Ehlotzky.RPP.2009,Piazza.RMP.2012,Zhang.PoP.2020,Fedotov.2022}.
One of the most important strong-field QED phenomena is the nonlinear Breit-Wheeler process (NBW)~\cite{Breit.PR.1934,Reiss.JMP.1962,Nikishov.JETP.1964,Yakovlev.JETP.1966} --- the production of an electron-positron pair from the interaction between a high-energy $\gamma$-photon and strong electromagnetic field.
High-power lasers are an ideal source of strong-fields, with field strengths $E_{0} \sim 10^{-3} E_{\text{S}}$ already achieved experimentally with PW-class systems~\cite{Yoon.Optica.2021}, where $E_{\text{S}} \sim 1.32 \times 10^{18}$~Vm$^{-1}$ is the Schwinger critical field of QED at which non-perturbative pair production occurs~\cite{sauter1931verhalten,heisenberg1936folgerungen,Schwinger.PR.1951}.

With high-power lasers supplying the strong-fields, one still requires a source of $\gamma$-photons for NBW experiments.
The source should ideally meet the following criteria: 
\begin{enumerate}[label=\roman*.]
    \item\label{i} \emph{High-energy} --- NBW is exponentially suppressed when the quantum nonlinearity parameter for a photon with momentum $l_{\mu}$, $\chi_{\gamma} = \sqrt{- (F_{\mu\nu} l^{\nu})^{2}}/(m c E_{\text{S}}) \ll 1$, becoming more probable as $\chi_{\gamma} \gtrsim 1$.
        If the field strength of the laser pulse is parameterised by the dimensionless intensity parameter, $\xi = e E_{0} \lambda_{\text{C}}/\hbar \omega_{0}$,\footnote{Here, $e$ is the electron charge, $E_{0}$ is the electric field strength, $\lambda_{\text{c}} = \hbar/mc$ is the Compton wavelength of an electron with mass $m$, and $\omega_{0}$ is the central frequency of the laser pulse.} this corresponds to a photon energy $\omega_{\gamma} \gtrsim m^{2}/(2 \omega_{0} \xi)$.
        Typical multi-PW laser facilities will operate with optical frequencies, $\omega_{0} \sim 1$eV, and field strengths $\xi \sim 10^{2} - 10^{3}$, requiring photons with energy in the MeV---GeV range.

        \vspace{-0.5em}

    \item\label{ii} \emph{Large numbers} --- the total number of generated electron-positron pairs, $\mcN_{e^{-}e^{+}}$, is directly proportional to the number of photons which collide with the laser, $\mcN_{e^{-}e^{+}} \propto \mcN_{\gamma}$.

        \vspace{-0.5em}

    \item\label{iii} \emph{Synchronised} --- multi-PW laser systems reach high peak power by compressing laser pulses to femtosecond (fs) durations.
        The photon source should be easily synchronised with the colliding pulse to ensure large numbers of photons pass through the spatio-temporal region of highest field strength.

        \vspace{-0.5em}

        \item\label{iv} \emph{Overlap} --- high intensities are achieved by focussing laser pulses to (typically) micron ($\mu\text{m}$) beam waists, $w_{0}$.
                The photon beam should have large spatial overlap with the laser focal spot to mitigate the impact of shot-to-shot fluctuations.

        \vspace{-0.5em}

    \item\label{v} \emph{Efficient} --- photon generation mechanism should efficiently convert the total input energy into a comparable total energy of photons.
\end{enumerate}
Different photon sources suitable for strong-field QED experiments have been proposed, which generally fall into two categories: electron-seeded or laser-driven (for a review see, e.g.~\cite{Gonoskov.RevModPhys.94.045001.2022} and references therein).

\begin{figure*}[t!!]
    \centering
    \includegraphics[width=0.65\linewidth]{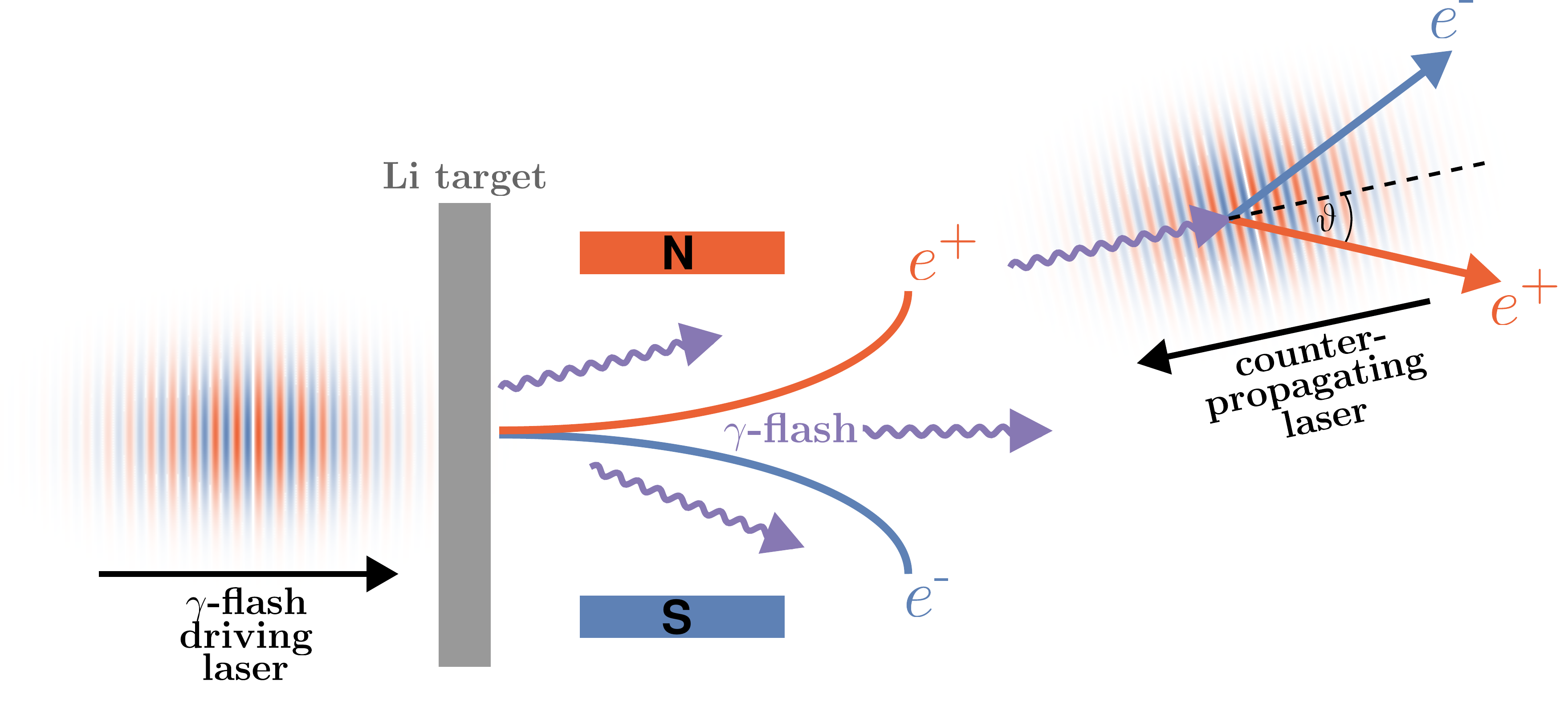}
    \caption{%
        Proposed experimental configuration.
        The total available laser energy, $E_{\text{total}}$, is split into two beams: $E_{\text{total}} = E_{\text{flash}} + E_{\text{pairs}}$.
        The beam with energy $E_{\text{flash}}$ is used to irradiate a solid lithium target, producing high-energy photons via the so-called $\gamma$-flash mechanism~\cite{Ridgers.PRL.2012,Nakamura.PRL.2012}.
        Charged secondary particles from the target are deflected to minimise background.
        The $\gamma$-flash photons propagate from the rear surface of the lithium target and collide with a counter-propagating secondary pulse, energy $E_{\text{pairs}}$, to produce electron-positron pairs via the nonlinear Breit-Wheeler mechanism~\cite{Breit.PR.1934,Reiss.JMP.1962,Nikishov.JETP.1964,Yakovlev.JETP.1966}.
        Positrons are emitted at an angle $\vartheta$ relative to the axis of the colliding counter-propagating laser.
    }
    \label{fig:Scheme}
\end{figure*}

Electron-seeded schemes generate $\gamma$-photons by colliding electron bunches with electromagnetic fields or high-$Z$ targets.
In the former case, photons can be produced in the perturbative regime, $\xi \ll 1$, via inverse Thomson/Compton scattering~\cite{Esarey.PRE.1993}, or in the non-perturbative regime, $\xi \gtrsim 1$, via nonlinear Compton scattering~\cite{Nikishov.JETP.1964,Brown.PR.1964,Goldman.PL.1964}.
The weak-field case, $\xi \ll 1$, produces radiation which is highly monochromatic and polarised~\cite{Hartemann.PRSTAB.2005,King.PRA.2020.b,Tang.PLB.2020}, but requires high-density electron bunches of GeV energy to produce the significant numbers of MeV photons~\cite{Hartemann.PRSTAB.2005,Chen.PRL.2013} required for NBW.
In the nonlinear regime, $\xi \gtrsim 1$, electron bunches of comparatively lower density can be used to generate high-brightness photon beams with energy comparable to the initial energy of the electrons (see e.g.~\cite{Sarri.PRL.2014,Yan.NP.2017}), due to the high field-strengths causing significant portions of the electron energy to be radiated.
Alternatively, electron beams can be collided with high-Z targets to produce photons via bremsstrahlung~\cite{Galy_2007,Ledingham_2010}, where the maximum photon energy is again comparable to the initial electron energy.
This scheme has shown promise for the study of NBW~\cite{Blackburn.PPCF.2018}, and is being considered as the primary source of photons in several experimental proposals (see e.g.~\cite{Abramowicz.EPJST.2021,Mercuri-Baron.NJP.2021,Eckey.PRA.2021,Golub.PRD.2022}).
Novel schemes have also been proposed utilising electron beam-multifoil collisions~\cite{Sampath.PRL.2021} and high-density electron bunch collisions with solid targets~\cite{Matheron:2022ajk} to produce $\gamma$-photons with high conversion efficiency.

The key limiting factor of each of these approaches is the initial step of producing high-energy, high-density, electron beams.
This must be achieved using conventional RF accelerators or laser-wakefield acceleration (LWFA)~\cite{Tajima.PRL.1979,Esarey.RMP.2009,Gonsalves.PRL.2019}.
Conventionally accelerated electron beams were utilised in the first experimental demonstration of pair production in the weak field regime ($\xi \ll 1$)~\cite{Bula.PRL.1996,Bamber.PRD.1999}, and will be used in the upcoming LUXE campaign to explore pair production in the transition regime ($\xi \gtrsim 1$)~\cite{Abramowicz.EPJST.2021}.
However, currently no facility exists which hosts \emph{both} a conventional accelerator and multi-PW laser system, putting the non-perturbative multi-photon regime of NBW ($\xi \gg 1$) out of reach with this approach.
Therefore, LWFA will be the primary mechanism for producing electron bunches at multi-PW laser facilities.
A typical photon source using LWFA electrons will require multiple stages.
First, the electrons are produced through a single/multi-stage acceleration scheme with an initial laser pulse colliding with an under-dense plasma.
Secondly, depending on the properties of the electron beam such as its transverse size and divergence, these will need to be focussed/columnated to achieve higher densities and mitigate undesirable features in the produced photons (see e.g.~\cite{Blackburn.PPCF.2018}).
Finally, the electrons will generate photons via one of the mechanisms outlined above, requiring either another laser pulse or strong-field source, or collision with a high-$Z$ target.
At each stage in the photon generation scheme, nonlinear plasma effects, shot-to-shot fluctuations in laser parameters and/or electron beam properties, and the spatio-temporal size of the produced photon beams can make synchronisation with another colliding laser pulse extremely challenging.
Furthermore, the conversion efficiency between the initial laser energy and the total energy of the produced photons can be extremely low.

Instead of using electrons to generate $\gamma$-photons, one can instead hope to produce them more directly using a laser-driven approach.
While many of the electron-seeded schemes described above would require multi-stage experimental configurations and access to high-energy electron sources such as conventional accelerators, laser-driven $\gamma$-photon generation can typically be achieved in a single stage, with the only requirement being access to a high-power laser.
A simple scheme which uses high-power lasers to irradiate solid targets is the so-called $\gamma$-flash mechanism~\cite{Ridgers.PRL.2012,Nakamura.PRL.2012}.
The $\gamma$-flash mechanism meets all of the desired properties \textrm{i}---\textrm{v} outlined above, producing large numbers of MeV---GeV photons with very high conversion efficiency between the laser energy and the energy of the produced photons~\cite{Brady.PoP.2014,Li.PRL.2015,Zhu_2015,Stark.PRL.2016,Lezhnin.PoP.2018,Gu.CommPhys.2018,Huang_2019,Hadjisolomou.JoPP.2021,Hadjisolomou.2022,Zhang_2015,Grismayer.PoP.2016,Zhu.NatCom.2016,Vranic_2016,Gonoskov.PRX.2017,Gong.PRE.2017,Efimenkp.SciRep.2018}. 
Furthermore, the use of only a single laser-driven stage to generate the $\gamma$-flash, coupled with the short duration and large transverse size of the photon beam, makes synchronisation to, and overlap with, a secondary laser pulse particularly simple.

In this paper we investigate the feasibility of using the $\gamma$-flash mechanism for studying nonlinear Breit-Wheeler pair production through an extremely simple \emph{all-optical} two-stage configuration, demonstrated in \cref{fig:Scheme}.
The key physical parameter which limits the attainable on-target peak power at a high-power laser facility is the total available pulse energy, $E_{\text{total}}$.
Our scheme assumes $E_{\text{total}}$ is split between two laser pulses, $E_{\text{total}} = E_{\text{flash}} + E_{\text{pairs}}$.
A pulse of energy $E_{\text{flash}}$ is used to irradiate an over-dense plasma, chosen as solid lithium (Li), to drive $\gamma$-photon production through the $\gamma$-flash mechanism.
This produces a $\gamma$-photon beam with large numbers of MeV---GeV photons, which propagate out from the rear surface of the target.
A second laser pulse, with energy $E_{\text{pairs}}$, collides head-on with the $\gamma$-photons at an interaction distance, $d$, from the target rear surface to produce electron-positron pairs through NBW.

The paper is structured as follows.
Firstly, in \cref{sec:Spectrum} we discuss the angular and spectral properties of photons generated through the $\gamma$-flash mechanism.
The spectra are produced using the particle-in-cell (PIC) code EPOCH~\cite{Arber_2015}.
In \cref{sec:BW} we summarise theoretical aspects of NBW, giving expressions for the differential probability of pair production from the collision of a photon and a linearly polarised plane wave pulse with Gaussian temporal envelope.
In \cref{sec:PairsFromFlash} numerical results are presented for the total number of electron-positron pairs produced through the interaction of $\gamma$-flash photons with high-power laser pulses.
We consider three different cases of total available laser pulse energy, $E_{\text{total}}$, relevant for current and next generation laser facilities, and discuss the optimal partitioning of this energy into $E_{\text{flash}}$ and $E_{\text{pairs}}$ to maximise the overall pair yield. 
We also discuss the energy and angular properties of the produced positrons.
Finally, in \cref{sec:Summary}, we summarise our key findings and discuss future steps for refining and optimising our approach.  
Throughout the rest of the paper we work in natural units, $\hbar = c = 1$, unless otherwise specified and use the shorthand notation $a^{\mu} b_{\mu} \equiv a \cdot b$ and $b \cdot b \equiv b^{2}$ with metric tensor, $g_{\mu\nu} = \text{diag}(+1,-1,-1,-1)$.

\section{$\gamma$-flash photon spectrum \label{sec:Spectrum}}

At low intensities, \emph{over-dense plasmas} --- where the electron density $n_{e} > n_{\text{cr}}$ with $n_{\text{cr}} = \epsilon_{0} m_{e} \omega_{0}^{2}/e^{2}$ the critical density for frequency $\omega_{0}$ --- are opaque to laser light.
Irradiating an over-dense plasma with a laser of sufficiently high intensity induces relativistic transparency, allowing the laser to propagate into the plasma and drive electron motion.
A dense ``QED plasma'' is produced where there is an interplay between field-induced QED phenomena and collective plasma effects~\cite{Ridgers.PRL.2012,Nakamura.PRL.2012}.
Copious numbers of high-energy photons are generated by charged particles in the QED plasma by a combination of nonlinear Compton scattering~\cite{Ridgers.PRL.2012,Nakamura.PRL.2012} and bremsstrahlung~\cite{Galy_2007,Ledingham_2010}, in a mechanism often referred to as a ``$\gamma$-flash''.
Numerical studies primarily utilising particle-in-cell (PIC) codes~\cite{Ridgers.PRL.2012,Nakamura.PRL.2012} have demonstrated that the laser-to-photon energy conversion efficiency, $\kappa_{\gamma}$, for the $\gamma$-flash mechanism can be as large as several tens of percent for single laser~\cite{Brady.PoP.2014,Li.PRL.2015,Zhu_2015,Stark.PRL.2016,Lezhnin.PoP.2018,Gu.CommPhys.2018,Huang_2019,Hadjisolomou.JoPP.2021,Hadjisolomou.2022}, dual laser~\cite{Zhang_2015,Grismayer.PoP.2016,Zhu.NatCom.2016} and multi-laser configurations~\cite{Vranic_2016,Gonoskov.PRX.2017,Gong.PRE.2017,Efimenkp.SciRep.2018}.

\begin{table}[t!!]
    \centering
    \begin{tabular}{| c | c | c | c | c | c | c |}
        \hline
        \multicolumn{3}{|c|}{Driving pulse $E_{\text{flash}}$} \\
        \hline
        $\tau_{\text{FWHM}}$ [fs] & $\lambda_{0}$ [$\mu$m] & $w_{0}$ [$\mu$m] \\
        \hline
        17 & 0.815 & 1.86 \\
        \hline
    \end{tabular}
    \caption{
        Constant laser parameters for $\gamma$-flash driving laser with energy $E_{\text{flash}}$.
    }
    \label{tab:Flash}
\end{table}

The first stage of our setup involves irradiating an over-dense plasma with an intense laser pulse to generate high-energy photons.
The plasma is taken to be a solid lithium (Li) target, density $n_{e} \approx 1.39 \times 10^{29}$~m$^{-3}$, with a diameter $12~\mu$m and thickness $10~\mu$m.
The choice of target is twofold.
Firstly, thin low-Z targets are known to reduce secondary particle production by photons in the material, compared to thicker and/or higher-Z targets (see e.g.~\cite{Vyskocil.PPCF.2018,Kotelanty.PhysRevResearch.4.023124.2022,Chintalwad.PRE.2022}).
Secondly, Li has previously been shown to optimise the laser-to-photon energy conversion efficiency with metallic targets~\cite{Hadjisolomou.2022}.
To increase the efficiency of photon generation the target is first irradiated with a long pre-pulse which generates a conical channel and has a similar effect to using targets fabricated with cone structures, see e.g.~\cite{Chintalwad.PRE.2022,Budriga:20,Badziak.PoP.2012.,Busold.PRSTAB.2013,Hadjisolomou.2022}.
The three-dimensional electron number density for the structured target is reproduced from a publicly available dataset~\cite{Tsygvintsev.RHD.2022} which calculates the effect of the pre-pulse using radiation hydrodynamic (RHD) simulations.
This data is then used as the initial conditions for the 3D-PIC simulations which model photon generation using the code EPOCH~\cite{Arber_2015} compiled with the Higuera-Cary~\cite{Higuera.PoP.2017}, bremsstrahlung and photons~\cite{Ridgers.JCompPhys.2014} directives enabled.

\begin{figure}[b!!]
    \centering
    \includegraphics[width=0.99\linewidth]{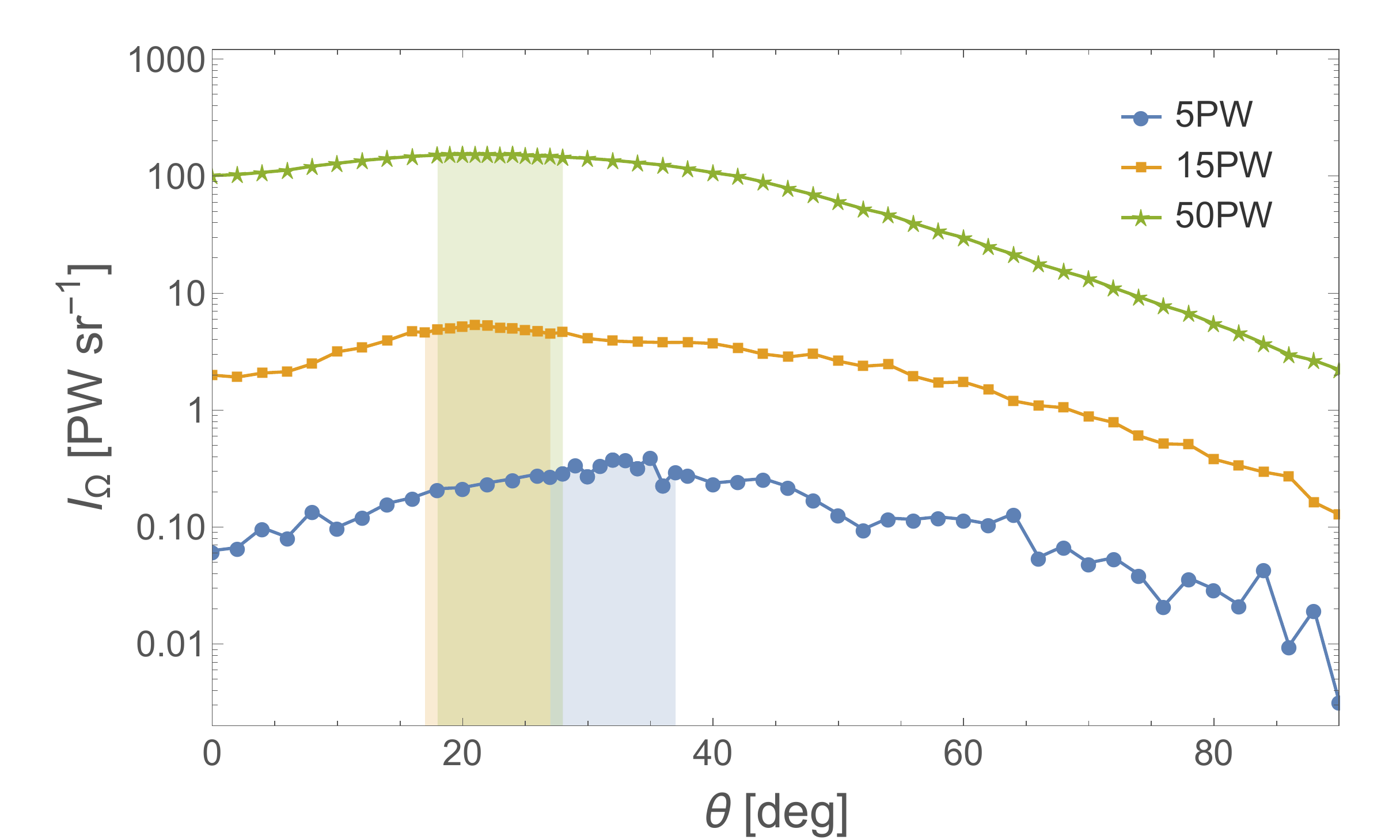}
    \caption{%
        Angular distribution of radiant intensity of $\gamma$-flash photons along laser polarisation axis, averaged over the two-lobe angular structure. 
        $\gamma$-flash produced by interaction with Li solid target and focussed laser pulses with peak powers 5~PW (blue, circles), 15~PW (orange, squares) and 50~PW (green, stars).
        Shaded region denotes $\beta = 10~\text{deg}$ full-angle divergence which maximises the radiant intensity.
    }
    \label{fig:AngleSpectra}
\end{figure}

We consider the gamma photon spectra generated from lasers with different values of the laser energy, $E_{\text{flash}}$.
In each case the pulse is a linearly polarised laser with full width half maximum (FWHM) duration of 17~fs focussed at normal incidence on the target with beam waist $w_{0} \sim 1.86~\mu$m. 
The central laser wavelength is $\lambda_{0} = 0.815~\mu$m, typical of Ti:sapphire laser systems~\cite{Spence.OL.1991} used at many current and next-generation facilities (see e.g.~\cite{Bromage.HPLSE.2021,Gan.2021,Papadopoulos.HPLSE.2016,Gales.RPP.2018,Sung.OL.2017,Yoon.Optica.2021}).
These parameters are summarised in \cref{tab:Flash}.

Photons are emitted in a symmetric double-lobe pattern due to the transverse motion of electrons in the plasma (see e.g.~\cite{Nakamura.PRL.2012,Ji.PhysRevLett.112.145003.2014,Duff.PPCF.2018,Vyskocil.PPCF.2020,Hadjisolomou.JoPP.2021,Hadjisolomou.2022}).
The angular distribution of the radiant intensity of the $\gamma$-flash photons from the rear surface of the Li target for three different laser energies, $E_{\text{flash}} = (85,255,850)$~J, is shown in~\cref{fig:AngleSpectra}.
For the constant parameters \cref{tab:Flash}, these correspond to laser powers, $P_{\text{flash}} = (5,15,50)$~PW, and approximate intensities $I_{\text{flash}} \sim 5 \times 10^{22} - 5 \times 10^{23}$~Wcm$^{-2}$ (or equivalently, $\xi_{\text{flash}} \sim 150 - 475$).
An angle of $\theta = 0\text{deg}$ corresponds to emission along the laser beam axis.
The radiant intensity of the $\gamma$-flash photons has been averaged over the symmetric double-lobe emission pattern.
\begin{figure}[t!!]
    \centering
    \includegraphics[width=0.99\linewidth]{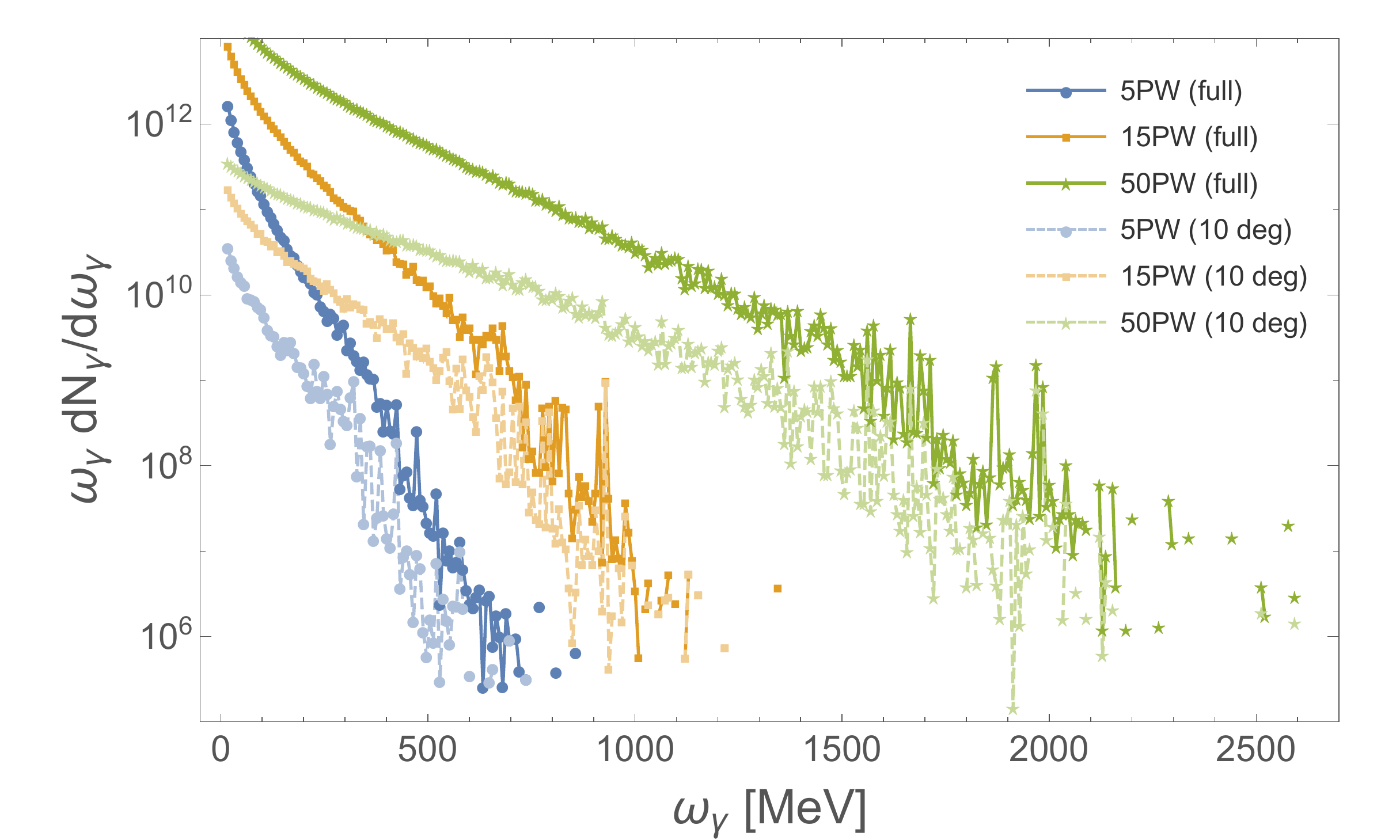}
    \caption{%
        $\gamma$-flash energy spectrum $\omega_{\gamma} \ud \mcN_{\gamma}/d\omega_{\gamma}$ generated by lasers with peak powers 5~PW (blue, circles), 15~PW (orange, squares) and 50~PW (green, stars).
        Dark solid lines show energy spectrum over all emission angles.
        Light dashed lines show energy spectrum over peak with full-angle divergence $\beta = 10~\text{deg}$, corresponding respective shaded regions in \cref{fig:AngleSpectra}.
    }
    \label{fig:Spectra}
\end{figure}

By colliding the produced photons with a secondary laser pulse, they will act as seed photons for NBW.
This process requires a large flux of high-energy photons and so to maximise the number of pairs produced the secondary laser should be focussed to the region of highest radiant intensity.
The energy spectra which will be used in the calculation of NBW in \cref{sec:PairsFromFlash} will correspond to the $\gamma$-flash photons within the full-angle divergence of $\beta = 10~\text{deg}$ which maximises the radiant intensity.
The appropriate region for each of the different laser powers is shown as the shaded portions of~\cref{fig:AngleSpectra}.
These are found to be centered around $\theta = (32,22,23)\text{deg}$, for respectively $P_{\text{flash}} = (5,15,50)$~PW.

\begin{figure}[t!!]
    \centering
    \includegraphics[width=0.99\linewidth]{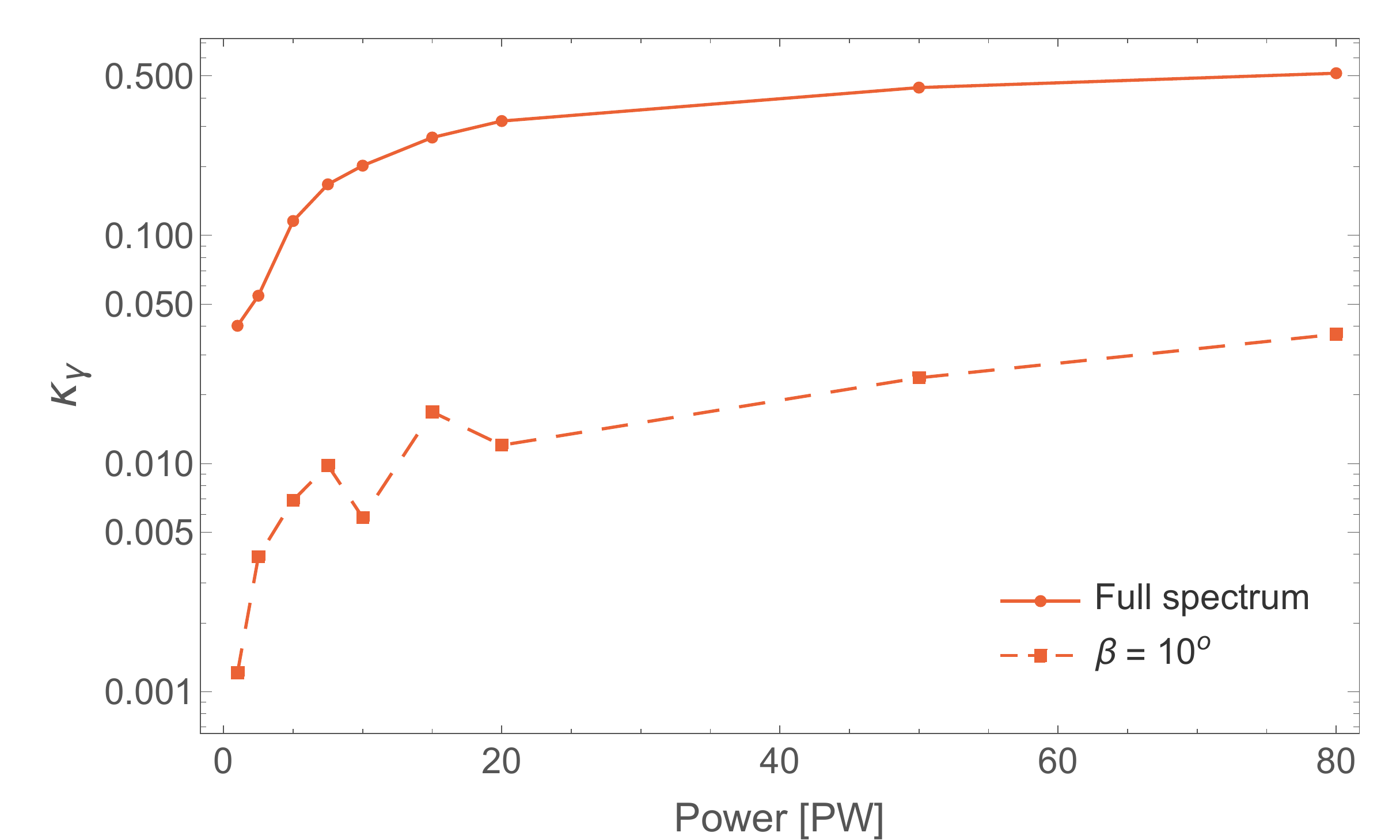}
    \caption{%
        Laser-to-photon energy conversion efficiency $\kappa_{\gamma}$ versus laser peak power for the full spectrum (solid) and for the photons emitted within the full-angle divergence $\beta = 10~\text{deg}$ (dashed).}
    \label{fig:Conversion}
\end{figure}

The differential energy spectra, $\omega_{\gamma} \ud \mcN_{\gamma}(\omega_{\gamma})/\ud \omega_{\gamma}$, (energy $\omega_{\gamma}$, differential number of photons $\ud\mcN_{\gamma}(\omega_{\gamma})/\ud\omega_{\gamma}$) of the $\gamma$-flash photons are shown in \cref{fig:Spectra}.
Dark solid colour curves give the full angular spectra, with the light dashed curves showing the spectra corresponding photons within the $\beta = 10~\text{deg}$ full-angle divergence shaded regions of \cref{fig:AngleSpectra}.
The photons within the $10~\text{deg}$ full-angle divergence account for a large number of the highest energy photons in the $\gamma$-flash spectrum, with lower-energy photons primarily filtered out.
Calculating the laser-to-photon energy conversion efficiency for the full angular spectra, see \cref{fig:Conversion}, we reaffirm previous analyses demonstrating very high efficiencies for the full angular spectrum~\cite{Ridgers.PRL.2012,Nakamura.PRL.2012,Brady.PoP.2014,Li.PRL.2015,Zhu_2015,Stark.PRL.2016,Lezhnin.PoP.2018,Gu.CommPhys.2018,Huang_2019,Hadjisolomou.JoPP.2021,Hadjisolomou.2022,Zhang_2015,Grismayer.PoP.2016,Zhu.NatCom.2016,Vranic_2016,Gonoskov.PRX.2017,Gong.PRE.2017,Efimenkp.SciRep.2018}.
The proportion of laser energy converted into the $\beta = 10~\text{deg}$ full-angle divergence is also shown.

\section{Nonlinear Breit-Wheeler pair production \label{sec:BW}}

The theory of NBW is by now very well established, and we discuss only the key features.
For more details and references see~\cite{Ritus.1985,Ehlotzky.RPP.2009,Piazza.RMP.2012,Zhang.PoP.2020,Fedotov.2022}.
The \emph{linear} Breit-Wheeler process is the production of an electron-positron pair from two \emph{real}\footnote{For the case of pair production from \emph{virtual} photons, typically known as the Bethe-Heitler process, see~\cite{Bethe.ProcRSoc.1934}.} high-energy $\gamma$-photons~\cite{Breit.PR.1934}.
To produce a pair, the total energy of the $\gamma$-photons must exceed the minimum threshold $\omega_{1} + \omega_{2} \ge 2m$.
Alongside the smallness of the corresponding cross-section, this has made linear Breit-Wheeler extremely difficult to observe experimentally~\cite{STAR.PRL.2021,He.ComPhys.2021}.
NBW is a multi-photon generalisation, where pairs are produced from the interaction of a \emph{single} high-energy gamma-photon with a \emph{large number} of (typically) low-energy photons which are sourced from a strong electromagnetic field~\cite{Reiss.JMP.1962,Nikishov.JETP.1964,Yakovlev.JETP.1966}.
If the field is a plane wave pulse with dimensionless field-strength $\xi$, wavevector $k_{\mu}$, then NBW obeys the 4-momentum conservation relation $l_{\mu} + \nu k_{\mu} = p_{\mu} + q_{\mu}$, where $l_{\mu}~(l^{2} = 0)$ is the momentum of the high-energy $\gamma$-photon, $p_{\mu}$ and $q_{\mu}~(p^{2} = q^{2} = m^{2})$ are the momenta of the produced electron and positron, and $\nu$ is the proportion of energy-momentum absorbed from the plane wave background.
The threshold for NBW can then be defined locally as $\nu \ge (2 + \xi^{2}(\phi))/\eta_{\gamma}$~\cite{Heinzl.PRA.2020}, where $\xi(\phi)$ is the local value of the intensity parameter and 
\begin{align}\label{eqn:eta}
    \eta_{\gamma} = \frac{k \cdot l}{m^{2}}
    \,,
\end{align}
is the normalised momentum of the $\gamma$-photon along the direction of the colliding plane wave field.
To overcome this threshold one must have both of high-energy $\gamma$-photons and strong electromagnetic fields.

We consider the collision of photons with a linearly polarised plane wave field with Gaussian envelope, 
\begin{align}\label{eqn:GaugePotential}
    a_{\mu}(\phi)
    =
    m \xi \epsilon_{\mu} \cos\phi
    e^{- 4 \ln(2) \frac{\phi^{2}}{\Phi_{\text{FWHM}}^{2}}}
    \,,
\end{align}
where $\phi$ is the phase, $\epsilon_{\mu}$ denotes the polarisation direction, and $\xi$ is the dimensionless measure of the laser field strength~\cite{Heinzl.OC.2009}.
The full-width-half-maximum (FWHM) phase duration of the pulse, $\Phi_{\text{FWHM}}$, is related to the temporal FWHM via $\tau_{\text{FWHM}} = \Phi_{\text{FWHM}} \lambda_{0}/2c$, where $\lambda_{0}$ is the wavelength.
To reach the highest intensities, high-power lasers must be focussed.
However, it is known that strong focussing can be detrimental for NBW, see e.g.~\cite{Piazza.PRL.2016,Blackburn.PPCF.2018,Mercuri-Baron.NJP.2021,Golub.PRD.2022}.
To maximise the pair yield, the influence of transverse focussing effects should be minimised.
A focussed fundamental Gaussian beam propagating in the $z$-direction with focus at $z = 0$ has a radius of curvature at some position $z$ of $R(z) = z (1 + (z_{R}/z)^{2})$, where $z_{R} = \pi w_{0}^{2}/\lambda_{0}$ is the Rayleigh length.
As the radius of curvature becomes larger, i.e. when $(z_{R}/z)^{2} \gg 1$, the wavefronts of the beam begin to look more like those of a plane wave.
This condition can be used to define a maximum length scale, $z_{\text{max}}$, over which the focussed pulse can be approximated by a plane wave.
A reasonable choice would be to consider an order of magnitude in $(z_{R}/z)^{2}$, i.e. $z_{\text{max}} = z_{R}/3$.
This would also keep the beam radius approximately constant for $z < z_{\text{max}}$, i.e. $w^{2}(z) = w_{0}^{2} (1 + (z/z_{R})^{2}) \approx w_{0}^{2}$, such that focussing effects become minimised.
Using $z_{\text{max}}$ to set an upper bound on the laser FWHM pulse length, $c \tau_{\text{FWHM}}$, gives a condition for the minimum beam waist, $w_{0}$, for which the plane wave model is valid,
\begin{align}\label{eqn:PWCondition}
    w_{0} [\text{$\mu$m}] > 0.535 \sqrt{\tau_{\text{FWHM}} [\text{fs}] \times \lambda_{0} [\text{$\mu$m}]}
    \,.
\end{align}
The wavelength and FWHM duration of the colliding pulse are chosen to match the parameters used in \cref{sec:Spectrum}: $\lambda_{0} = 0.815~\mu$m and  $\tau_{\text{FWHM}} = 17$~fs.
This sets a lower bound of $w_{0} > 1.99~\mu$m.
Therefore, for each pulse energy, $E_{\text{pairs}}$, the beam waist will be chosen as $w_{0} = 2.5~\mu$m, i.e. $w_{0} \simeq 3 \lambda_{0}$.\footnote{Blackburn \& Marklund~\cite{Blackburn.PPCF.2018} demonstrated that for short pulses ($\tau_{\text{FWHM}} < 20$~fs) and small collision angles focussing effects only contribute a small change to the total pair production probability for $2 < w_{0}/\lambda_{0} < 10$. This is in agreement with our simple approximations and choice of parameters.}
The colliding pulse parameters are summarised in \cref{tab:Pairs}.

\begin{table}[b!!]
    \centering
    \begin{tabular}{| c | c | c | c | c | c | c |}
        \hline
        \multicolumn{3}{|c|}{Colliding pulse $E_{\text{pairs}}$} \\
        \hline
        $\tau_{\text{FWHM}}$ [fs] & $\lambda_{0}$ [$\mu$m] & $w_{0}$ [$\mu$m] \\
        \hline
        17 & 0.815 & 2.5 \\
        \hline
    \end{tabular}
    \caption{
        Constant laser parameters for colliding pulse, energy $E_{\text{pairs}}$, which drives pair production via NBW.
    }
    \label{tab:Pairs}
\end{table}

The NBW $S$-matrix element for the production of an electron-positron pair with momenta $(p_{\mu},q_{\mu})$ and spins $(s,r)$ from the collision of a photon of momentum $l_{\mu}$ and polarisation $\varepsilon_{l}^{\mu}$ with a plane wave pulse is,
\begin{align}\label{eqn:SMatrix}
    S_{fi}
    =
    -
    i
    e
    \int \ud^{4} x
    e^{- i l \cdot x}
    \bar{\psi}^{(-)}_{p,s}(x)
    \slashed{\varepsilon}_{l}
    \psi^{(+)}_{q,r}(x)
    \,,
\end{align}
where $e$ is the electron charge  and the Volkov wavefunctions~\cite{Volkov.ZP.1935} for the produced electron-positron pair are, respectively,
\begin{align}\label{epn:ElectronOut}
    \bar{\psi}^{(-)}_{p,s}(x)
    =
    e^{
        +
        i
        p \cdot x
        +
        i
        \int^{\phi} \ud t
        \frac{2 p \cdot a(t) - a^{2}(t)}{2 k \cdot p}
    }
    \bar{u}_{p}^{s}
    \bigg(
        1
        -
        \frac{\slashed{k} \slashed{a}(\phi)}{2 k \cdot p}
    \bigg) 
    \,,
    \\ \label{eqn:PositronOut}
    \psi^{(+)}_{q,r}(x)
    =
    e^{
        + 
        i 
        q \cdot x
        -
        i
        \int^{\phi} \ud t
        \frac{2 q \cdot a(t) + a^{2}(t)}{2 k \cdot q}
    }
    \bigg(
        1
        -
        \frac{\slashed{k} \slashed{a}(\phi)}{2 k \cdot q}
    \bigg) 
    v_{q}^{r}
    \,.
\end{align}
$\bar{u}_{p}^{s}$ and $v_{q}^{r}$ are free-space Dirac spinors and for any 4-vector $b_{\mu}$: $\slashed{b} \equiv \gamma^{\mu} b_{\mu}$ with $\gamma^{\mu}$ the Dirac matrices.

Calculations are performed in lightfront coordinates, $x^{\mu} = (x^{\lcm},x^{\lcperp},x^{\lcp})$, where $x^{\pm} = t \pm z$ and $x^{\lcperp} = (x,y)$.
The plane wave propagates in the $z$-direction with wavevector $k_{\mu} = \omega_{0} (1,0,0,1)$ and the phase is defined as $\phi \equiv k \cdot x = \omega_{0} x^{\lcm}$.
A generic on-shell 4-momentum, $p_{\mu}$, is expressed in lightfront variables as $p_{\mu} = (p_{\lcm},p_{\lcperp},p_{\lcp})$, where $k \cdot p = 2 \omega_{0} p_{\lcp}$, $p_{\lcperp} = (p_{x},p_{y})$, and the remaining component is fixed by the on-shell condition, $p^{2} = m^{2}$, as $p_{\lcm} = (m^{2} + p_{\lcperp}^{2})/4 p_{\lcp}$.
Integrals over $(x^{\lcperp},x^{\lcp})$ in \cref{eqn:SMatrix} yield momentum conserving $\delta$-functions.
The differential probability can then be found by taking the squared modulus of $S_{fi}$, averaging/summing over initial/final spins, and integrating over the electron momenta, $p_{\mu}$.
We are specifically interested in the interaction of MeV---GeV photons with multi-PW, multi-cycle, laser pulses, where the dimensionless intensity parameter $\xi \gg 1$.
As such, we are within the regime of validity\footnote{For discussions of the regime of validity of the LCFA see~\cite{Ritus.1985,Baier.NPB.1989,Dinu.PRL.2016}, and for extensions/alternatives see~\cite{Ilderton.PRA.2019,Piazza.PRA.2019,Heinzl.PRA.2020,King.PRA.2020}.} of the locally-constant field approximation (LCFA),
which gives the angularly resolved differential probability,
\begin{align}\label{eqn:AngularLCFA}
    \frac{\ud^{3} \sfP_{\text{LCFA}}}{\ud r \ud \eta_{q} \ud \psi}
    &
    =
    \frac{\alpha r}{\pi \eta_{\gamma}^{2}}
    \int \ud \phi
    \Ai(\bar{z}(\phi))
    \nonumber\\
    &
    \times
    \bigg\{
        z(\phi)
        +
        \bigg[
            \frac{(\eta_{\gamma} - \eta_{q})^{2} + \eta_{q}^{2}}{\eta_{q} (\eta_{\gamma} - \eta_{q})}
        \bigg] 
        \bar{z}(\phi)
    \bigg\} 
    \,,
\end{align}
where $\alpha = e^{2}/4\pi$ is the fine-structure constant.
The argument of the Airy function, $\Ai(\bar{z}(\phi))$, is,
\begin{align}\label{eqn:barz}
    \bar{z}(\phi)
    =
    z(\phi)
    \Big[
        1 + r^{2} + \frac{|a_{\lcperp}(\phi)|^{2}}{m^{2}} + \frac{2 r |a_{\lcperp}(\phi)| \cos\psi}{m}
    \Big] 
    \,,
\end{align}
where,
\begin{align}\label{eqn:z}
    z(\phi)
    =
    &
    \bigg(
        \frac{1}{\chi_{\gamma}(\phi)}
        \frac{\eta_{\gamma}^{2}}{\eta_{q} (\eta_{\gamma} - \eta_{q})}
    \bigg)^{2/3}
    \,,
\end{align}
is defined in terms of the quantum nonlinearity parameter of the photon,
\begin{align}\label{eqn:chi}
    \chi_{\gamma}(\phi)
    =
    &
    \frac{\eta_{\gamma} |a_{\lcperp}^{\prime}(\phi)|}{m}
    \,,
\end{align}
where $a_{\lcperp}^{\prime}(\phi) \equiv \ud a_{\lcperp}(\phi)/\ud\phi$.

The probability is compactly parameterised by three parameters, $(\eta_{q},r,\psi)$, where,
\begin{align}\label{eqn:PositronEta}
    \eta_{q}
    =
    \frac{k \cdot q}{m^{2}}
    \,,
\end{align}
is the normalised momentum of the positron along the direction of the colliding plane wave, $r = |r_{\lcperp}|$, where,
\begin{align}\label{eqn:rperp}
    r_{\lcperp}
    =
    \frac{q_{\lcperp} - \frac{\eta_{q}}{\eta_{\gamma}}l_{\lcperp}}{m}
    \,,
\end{align}
is a measure of the positrons momentum in the plane perpendicular to the direction of the laser, and $\psi \in [0,2\pi)$ is the azimuthal emission angle in the perpendicular plane, i.e. we could also write $r_{\lcperp} = r \{\cos\psi,\sin\psi\}$.
When the perpendicular momentum of the photon can be neglected, $q_{\lcperp} \gg (\eta_{q}/\eta_{\gamma}) l_{\lcperp}$, and the energy of the produced positron $E_{q} \gg m$, then $r \approx (E_{q}/m) \sin\vartheta$ and $\eta_{q} \approx (\omega_{0} E_{q}/m^{2}) (1 + \cos\vartheta)$, where $\vartheta$ is the emission angle relative to the colliding laser propagation axis (chosen here as the $z$-axis).
Thus, for small emission angles $\vartheta \ll 1$, $r \approx E_{q} \vartheta / m$ and $\eta_{q} \approx 2 \omega_{0} E_{q}/m$, and we can readily interpret the pair $(r,\eta_{q})$ as a parameterisation of the positron's emission angle and energy.

Integrating~\cref{eqn:AngularLCFA} returns the total probability, $\sfP_{\text{LCFA}}$, for NBW by a single photon with momentum, $l_{\mu}$.
However, in certain regimes $\sfP_{\text{LCFA}}$ can exceed unity and its interpretation as a probability becomes ambiguous. 
This is due to higher-order loop effects being neglected.
These can be included by solving the Schwinger-Dyson equations to arrive at photon wavefunctions which demonstrate an exponential decay~\cite{Meuren.PRD.2015}.
One can then define a \emph{decay probability},
\begin{align}\label{eqn:Survival}
    \sfW
    =
    1
    -
    \exp\big(- \sfP_{\text{LCFA}}\big) 
    \,,
\end{align}
in which $\sfP_{\text{LCFA}}$ now has the interpretation as the decay exponent for a photon with momentum, $l_{\mu}$, propagating through the laser pulse (see e.g.~\cite{Meuren.PRD.2015,Podszus.PRD.2021,Mercuri-Baron.NJP.2021,Tamburini.PRD.2021}).
When $\sfP_{\text{LCFA}} \ll 1$, then $\sfW \approx \sfP_{\text{LCFA}}$, and $\sfP_{\text{LCFA}}$ can again be interpreted as a probability.

\section{Pairs from gamma flash \label{sec:PairsFromFlash}}

Photons are produced at the target rear surface and are spread in the angular plane (c.f. \cref{fig:AngleSpectra}).
As discussed in \cref{sec:Spectrum} we consider photons within the full-angle divergence of $\beta = 10~\text{deg}$ for which the radiant intensity of the photons is maximised.
These photons will propagate a distance, $d$, to the colliding laser focus, expanding from the rear target surface as a spherical shell.
The distance, $d$, will typically be of $\mcO(10\text{cm})$, and the spherical shell of photons within the full-angle divergence $\beta = 10~\text{deg}$ will have a large radius of curvature and transverse size relative to the colliding laser beam waist, $w_{0} \sim \mcO(\mu\text{m})$.
The beam of photons which collide with the pulse can therefore be well approximated as a flat disk propagating from the target point source and the perpendicular momentum of the photons can be neglected, i.e. $\l_{\lcperp} \approx 0$, such that the photons collide approximately head-on with the counter-propagating laser pulse.
After propagating a distance, $d$, the photon beam will have an area $A_{\text{flash}} = \pi d^{2} \tan^{2}(\beta/2)$.
The photons will then collide with a counter-propagating laser pulse with a focal spot area $A_{\text{laser}} = \pi w_{0}^{2}$, where $w_{0}$ is the beam waist.
Given a total number of photons, $\mcN_{\gamma}$, within the full-angle divergence, $\beta$, the total number of photons within the laser focal spot, focussed at the distance $d$, will be\footnote{The validity of \cref{eqn:PhotonsFocus} requires $A_{\text{flash}} > A_{\text{laser}}$, which for $\beta = 10~\text{deg}$ is satisfied when $d \gtrsim 11 w_{0}$.},
\begin{align}\label{eqn:PhotonsFocus}
    \frac{\mcN_{\gamma} w_{0}^{2}}{d^{2} \tan^{2}(\beta/2)}
    \,.
\end{align}
Then, the total number of pairs generated from the collision of the photon beam with the colliding laser is,
\begin{align}\label{eqn:NumberPairs}
    \mcN_{e^{-}e^{+}}
    =
    &
    \frac{w_{0}^{2}}{d^{2} \tan^{2}(\beta/2)}
    \int_{0}^{\infty} \ud \omega_{\gamma}
    \sfW(\omega_{\gamma})
    \frac{\ud \mcN_{\gamma}(\omega_{\gamma})}{\ud \omega_{\gamma}}
    \,,
\end{align}
where $\sfW(\omega_{\gamma})$ is defined by~\cref{eqn:Survival} and $\ud\mcN_{\gamma}(\omega_{\gamma})/\ud\omega_{\gamma}$ is the differential number of photons with energy $\omega_{\gamma}$.

At a high-power laser facility, the key constraining parameter is the deliverable pulse \emph{energy}.
Current and upcoming PW and multi-PW laser facilities typically have deliverable pulse energies of $E_{\text{total}} \sim 30-1500$~J, with durations $\tau_{\text{FWHM}} \sim 15-30$~fs~\cite{Danson.HPLSE.2019} , with notable exceptions such as the 10 PW laser system at ELI Beamlines~\cite{Weber.MRE.2017} which will have a long pulse duration of $150$~fs.
The proposed scheme, \cref{fig:Scheme}, takes the total available laser energy and splits this into two, $E_{\text{total}} = E_{\text{flash}} + E_{\text{pairs}}$, where the pulse with energy $E_{\text{flash}}$ drives the photon production via the $\gamma$-flash mechanism and the pulse with energy $E_{\text{pairs}}$ collides with those photons to produce pairs. 
We shall consider three different cases for the total available laser energy, $E_{\text{total}} = (170, 510, 1700)$~J, and further consider different ratios 
\begin{align}\label{eqn:Delta}
    \Delta 
    =
    \frac{E_{\text{flash}}}{E_{\text{pairs}}}
    \,,
\end{align}
to find the optimal splitting of the total available laser energy for producing pairs.
Both pulses have a FWHM duration $\tau_{\text{FWHM}} = 17$~fs, such that the total available \emph{power} for each case is $P_{\text{total}} = (10, 30, 100)$~PW.
Current state-of-the-art technology has fuelled the development of a number of 10~PW class laser facilities~\cite{Weber.MRE.2017,Gales.RPP.2018,Gan.2021,Bromage.HPLSE.2021,Papadopoulos.HPLSE.2016}, with future facilities (e.g.~\cite{Mukhin.QE.2021,Shen.PPCF.2018}) aiming to break the 100~PW peak power threshold (see also~\cite{Tajima.PRSTAB.2002,Li.SciRep.2021}.
Our considerations will therefore explore the feasibility of using $\gamma$-flash photons to observe nonlinear Breit-Wheeler pair production at current-, next-, and future-generation high-power laser facilities.

\begin{figure}[t!]
    \centering
    \includegraphics[width=0.99\linewidth]{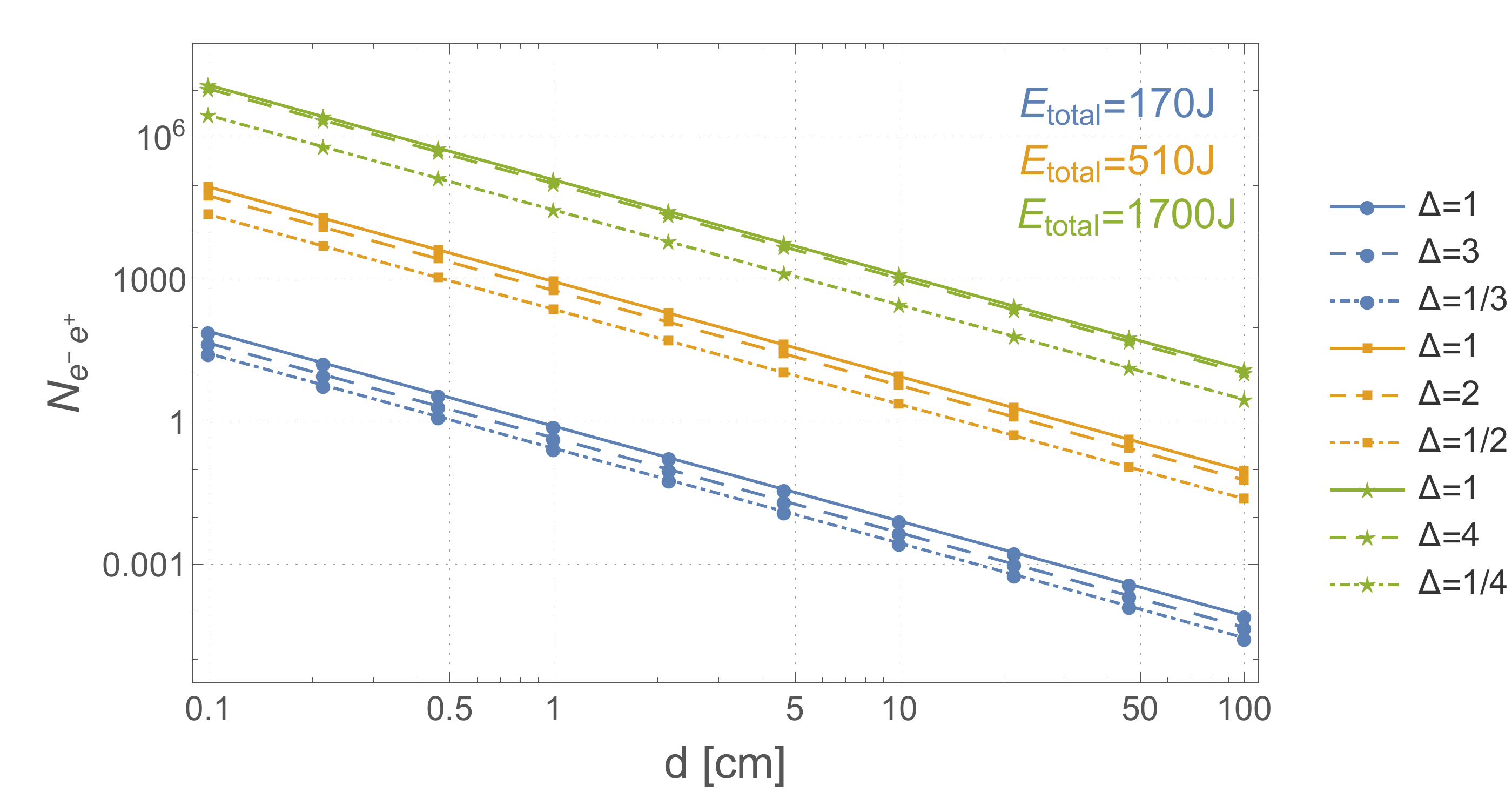}
    \caption{%
        Total number of produced pairs, $\mcN_{e^{-}e^{+}}$, as function of photon-laser interaction point distance, $d$, calculated with \cref{eqn:NumberPairs}.
        $E_{\text{total}} = 170$~J (blue, circles).
        $E_{\text{total}} = 510$~J (orange, squares).
        $E_{\text{total}} = 1700$~J (green, stars).
        Legend gives value of ratio $\Delta$ (c.f.~\cref{eqn:Delta}).
    }
    \label{fig:Total-Pairs}
\end{figure}

In \cref{fig:Total-Pairs} we plot the total number of pairs produced as the distance, $d$, is increased, using \cref{eqn:NumberPairs}.
The corresponding values of the peak intensities for the different total energies, $E_{\text{total}}$, and ratios, $\Delta$, used in \cref{fig:Total-Pairs} are outlined in \cref{tab:Data}.
Comparing the total number of pairs produced for different values of the splitting ratio, $\Delta$, at fixed $E_{\text{total}}$ and $d$, suggests that the number of pairs will be maximized when there is an equal split of the total energy into the beams which drive the $\gamma$-flash photon generation and pair production via NBW, i.e. when $\Delta = 1$.
For the lowest total laser energy, $E_{\text{total}} = 170$~J, we find $\mcN_{e^{-}e^{+}}^{\Delta=1}/\mcN_{e^{-}e^{+}}^{\Delta=3} \approx 1.8$ and $\mcN_{e^{-}e^{+}}^{\Delta=1}/\mcN_{e^{-}e^{+}}^{\Delta=1/3} \approx 2.9$.
At the intermediate energy, $E_{\text{total}} = 510$~J, we see a similar scaling between the ratios of the number of pairs with each value of $\Delta$, with $\mcN_{e^{-}e^{+}}^{\Delta=1}/\mcN_{e^{-}e^{+}}^{\Delta=2} \approx 1.6$ and $\mcN_{e^{-}e^{+}}^{\Delta=1}/\mcN_{e^{-}e^{+}}^{\Delta=1/2} \approx 3.9$.
Finally at the highest energy, $E_{\text{total}} = 1700$J, we find $\mcN_{e^{-}e^{+}}^{\Delta=1}/\mcN_{e^{-}e^{+}}^{\Delta=4} \approx 1.2$ and $\mcN_{e^{-}e^{+}}^{\Delta=1}/\mcN_{e^{-}e^{+}}^{\Delta=1/4} \approx 4.3$.

In the process of producing photons via the $\gamma$-flash mechanism, electrons and positrons are also created and emitted from the target rear surface (see e.g.~\cite{Ridgers.PRL.2012,Brady.PoP.2014,Gu.CommPhys.2018,Hadjisolomou.JoPP.2021,Zhu.NatCom.2016}).
The positrons are produced by high-energy photons inside the target, such that their energy and angular distribution are comparable to the emitted $\gamma$-flash photons, while electrons are produced both by pair production and by direct acceleration of target electrons by the irradiating laser pulse.
To minimise the background of charged particles an experiment would separate these from the photons using magnetic deflection, or other methods.
The photon propagation distance, $d$, therefore needs to be sufficiently large to allow for this background of particles to be filtered from the photons.
    For example, a permanent magnet of length, $L$, and field strength, $B$, can deflect an electron/positron propagating perpendicular to to the magnetic field with energy, $\mcE$, by an angle $\delta[\text{deg}] \simeq 0.17 L[\text{cm}] B[\text{T}] \mcE^{-1}[\text{GeV}]$.
    To ensure the background of charged particles at the focus of the colliding laser (energy $E_{\text{pairs}}$) is minimised, the deflection angle of particles propagating with the $\gamma$-flash photons should satisfy $\delta \gg \delta_{\text{min}}$ where $\delta_{\text{min}}[\text{deg}] \simeq 6 \times 10^{-3} w_{0}[\mu\text{m}]d^{-1}[\text{cm}]$.
    This corresponds to a magnet of length $L[\text{cm}] \gg 0.03 w_{0}[\mu\text{m}] \mcE[\text{GeV}] d^{-1}[\text{cm}] B^{-1}[T]$.
Considering a photon propagation distance of $d = 10$~cm, deflecting a positron, energy $\mcE = 3$~GeV, with a magnet of field strength $B = 1$~T away from the colliding laser focus of $w_{0} = 2.5~\mu$m would require $L \gg 0.02$~cm. 
A large number of background particles could therefore be removed with, for example, a $L = 5$~cm neodymium magnet.

\begin{table}[t!!]
    \centering
    \begin{tabular}{| c | c | c | c | c | c | c |}
        \hline
        $E_{\text{total}}$ [J] & $\Delta$ & $\xi_{\text{flash}}$ & $\xi_{\text{pairs}}$ & $I_{\text{flash}}$ [Wcm$^{-2}$] & $I_{\text{pairs}}$ [Wcm$^{-2}$] \\
        \hline
        170 & 1 & 149 & 111 & $4.6 \times 10^{22}$ & $2.5 \times 10^{22}$ \\
        - & 3 & 183 & 79 & $6.9 \times 10^{22}$ & $1.27 \times 10^{22}$ \\
        - & 1/3 & 106 & 136 & $2.3 \times 10^{22}$ & $3.8 \times 10^{22}$ \\
        \hline
        510 & 1 & 259 & 193 & $1.4 \times 10^{23}$ & $7.6 \times 10^{22}$ \\
        - & 2 & 299 & 157 & $1.84 \times 10^{23}$ & $5.1 \times 10^{22}$ \\
        - & 1/2 & 211 & 222 & $9.2 \times 10^{22}$ & $1.0 \times 10^{23}$ \\
        \hline
        1700 & 1 & 473 & 352 & $4.6 \times 10^{23}$ & $2.5 \times 10^{23}$ \\
        - & 4 & 598 & 222 & $7.36 \times 10^{23}$ & $1.0 \times 10^{23}$ \\
        - & 1/4 & 299 & 445 &  $1.84 \times 10^{23}$ & $4.0 \times 10^{23}$ \\
        \hline
    \end{tabular}
    \caption{
        Peak intensity parameters used to calculate total number of electron-positron pairs in~\cref{fig:Total-Pairs}.
        The constant parameters for the driving and colliding laser pulses are given in \cref{tab:Flash,tab:Pairs}, respectively.
        Intensity of each pulse given both in terms of dimensionless intensity parameters, $\xi_{\text{flash}}$ and $\xi_{\text{pairs}}$, and equivalent power per unit area, $I_{\text{flash}}$ and $I_{\text{pairs}}$.
    }
    \label{tab:Data}
\end{table}

With the colliding laser focus at $d = 10$~cm and using the optimal beam splitting ratio of $\Delta = 1$, the total number of pairs produced per shot for the different energies, $E_{\text{total}} = (170,510,1700)$~J, is $\mcN_{e^{-}e^{+}} \sim (0.01,10,1300)$, respectively.
A typical high-power laser has a repetition rate of the order of $0.1-10$~Hz~\cite{Danson.HPLSE.2019}.
This means that with an interaction distance $d = 10$~cm and nominal repetition rate of 0.1~Hz, an experiment could expect to produce approximately $\sim 5$ pairs/hour\footnote{
    This is comparable with the estimated number of NBW pairs which will be produced per hour at a proposed experiment with the CALA laser~\cite{Salgado.NJP.2021,Golub.PRD.2022}, which will use LWFA electrons to generate photon production via bremsstrahlung, c.f.~\cref{sec:Intro}.
} with the $10$~PW equivalent system, increasing to $\sim 10^{5}$ pairs/hour with $100$~PW.
If the interaction distance could be further reduced to $d=1$~cm, the number of pairs/shot increases substantially to, respectively, $\mcN_{e^{-}e^{+}} \sim (1, 1000, 10^{5})$ pairs/shot.

\begin{figure}[t!!]
    \centering
    \includegraphics[width=0.80\linewidth]{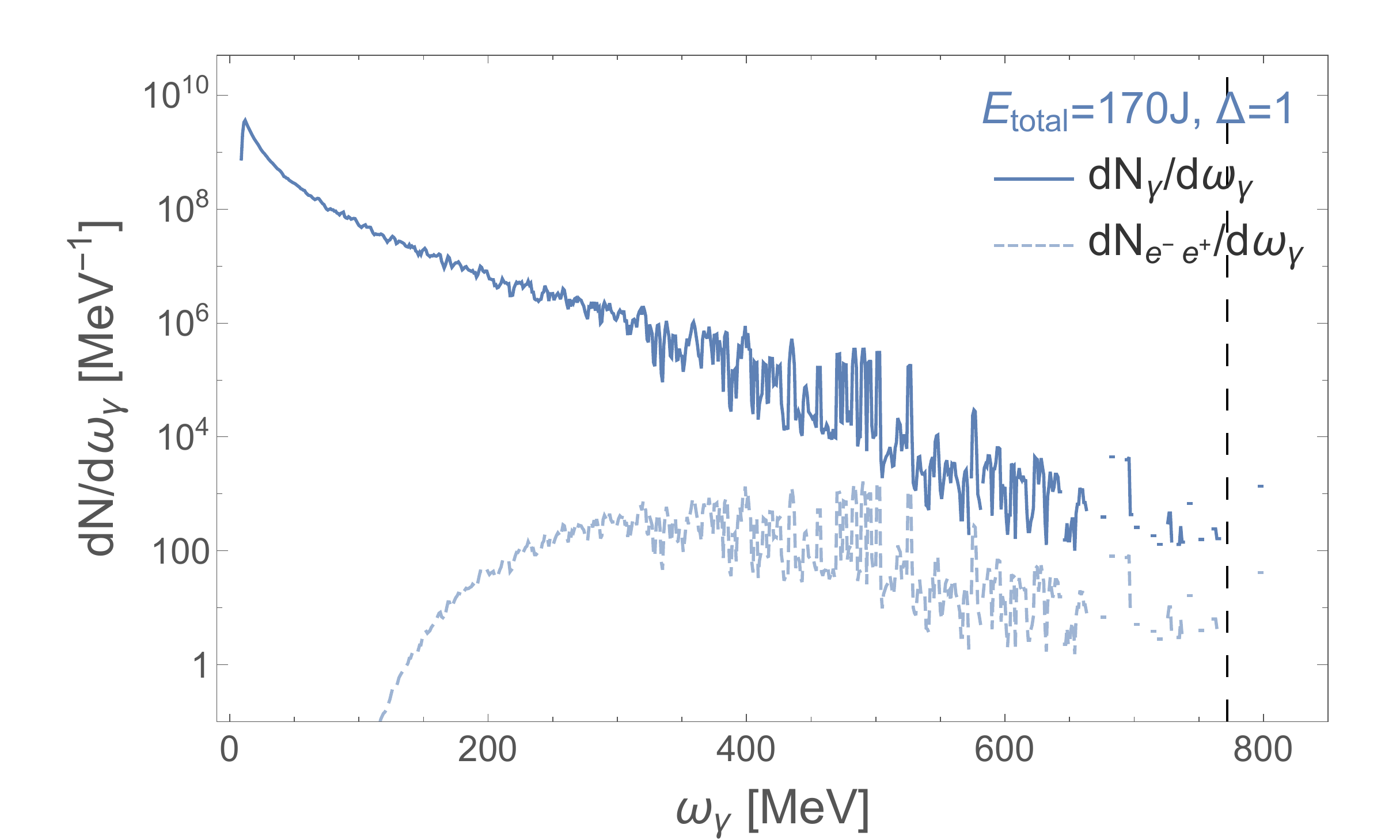}

    \includegraphics[width=0.80\linewidth]{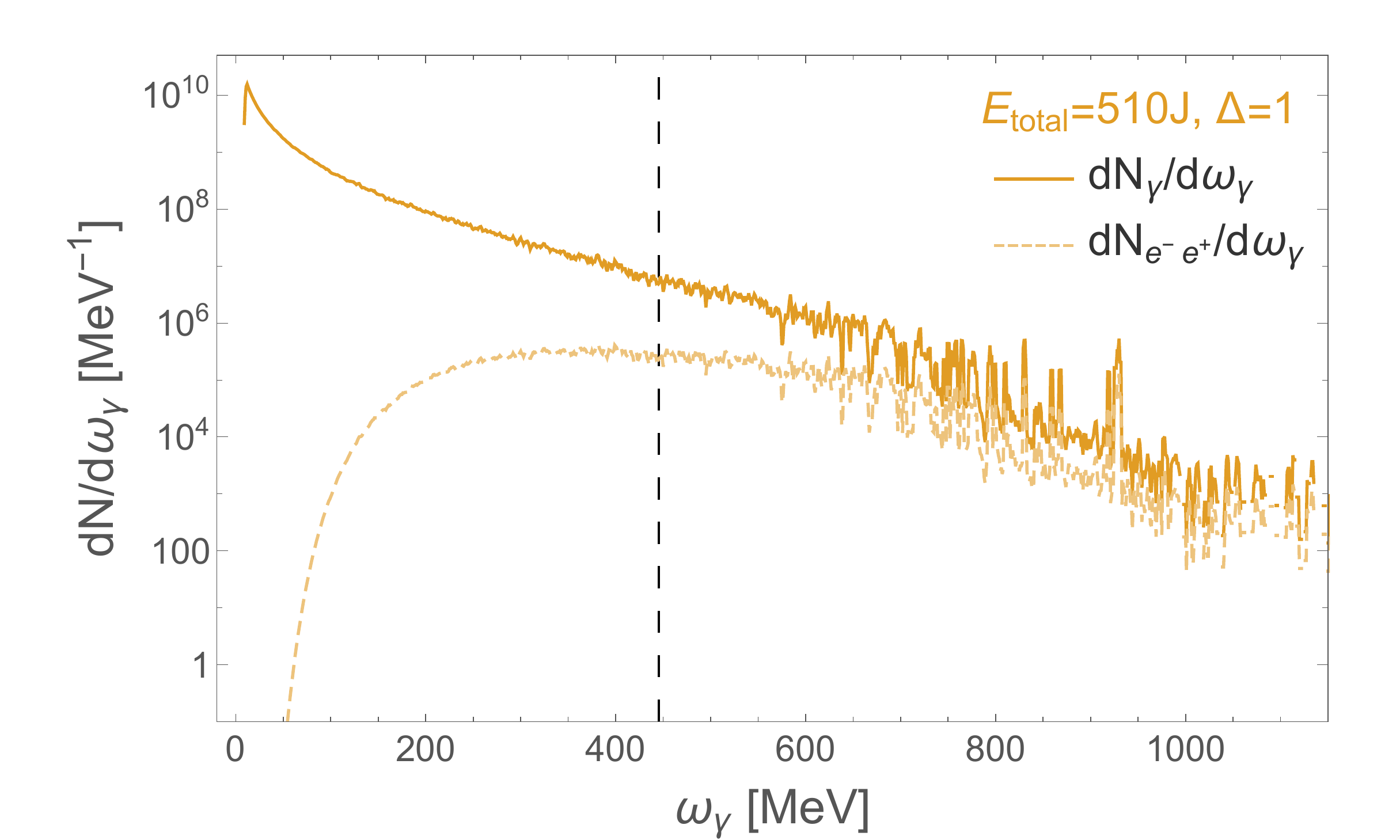}

    \includegraphics[width=0.80\linewidth]{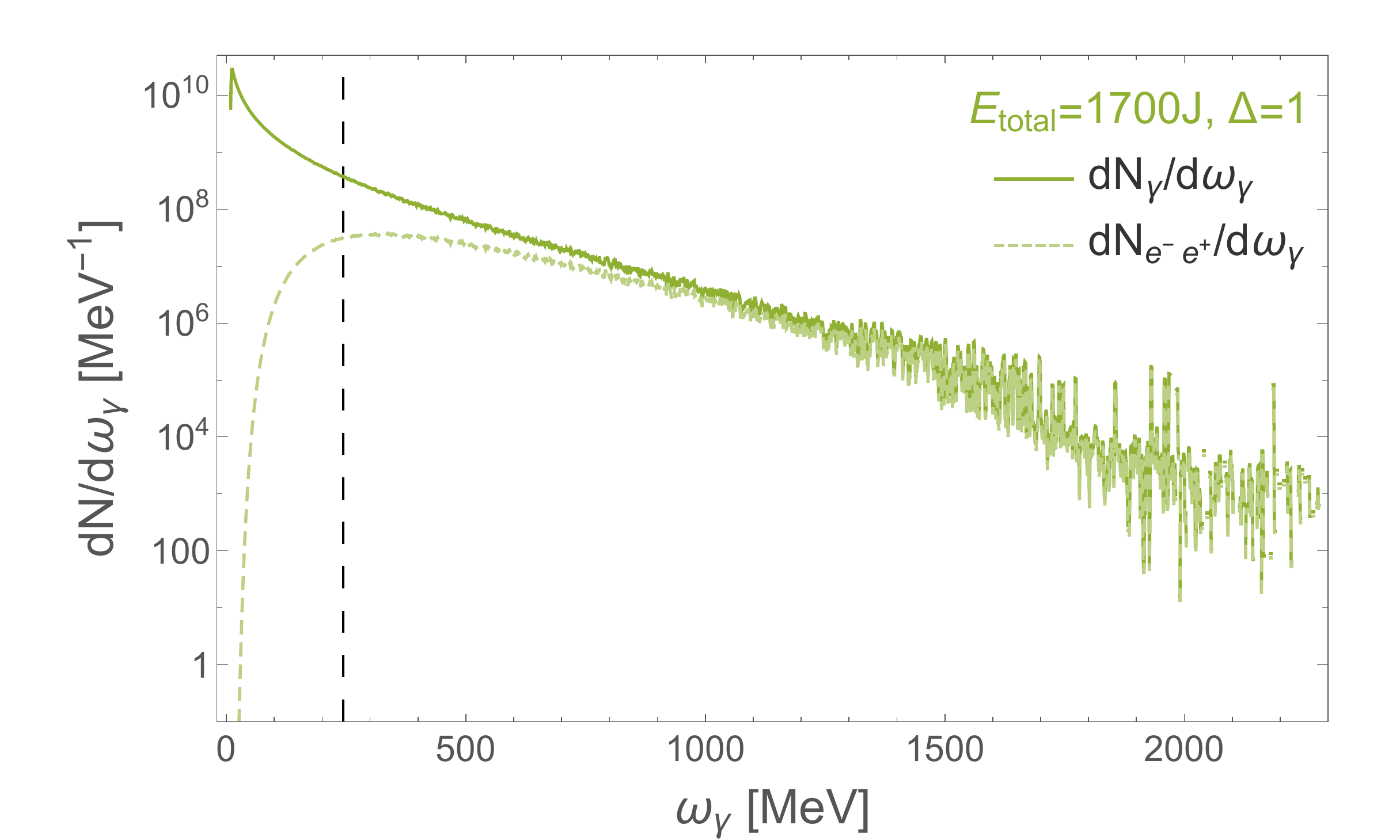}
    \caption{%
        Comparison of initial photon spectrum $\ud \mcN_{\gamma}/\ud \omega_{\gamma}$ (solid) with the differential number of pairs $\ud \mcN_{e^{-}e^{+}}/\ud \omega_{\gamma}$ (dashed).
        \emph{Top:} $E_{\text{total}} = 170$~J, $\Delta=1$.
        \emph{Middle:} $E_{\text{total}} = 510$~J, $\Delta=1$.                
        \emph{Bottom:} $E_{\text{total}} = 1700$~J, $\Delta=1$.
        Black dashed lines denote value of $\omega_{\gamma}$ for which $\text{max}[\chi_{\gamma}]=\eta_{\gamma}\xi_{\text{pairs}}=1$.
    }
    \label{fig:Spectra-Pairs-omega}
\end{figure}

For the three total laser energies, $E_{\text{total}}$, with the optimal splitting, $\Delta =1$, \cref{fig:Spectra-Pairs-omega} compares the differential number of pairs produced, $\ud \mcN_{e^{-}e^{+}}/\ud\omega_{\gamma}$ (light, dashed), with the differential number of photons which interact with the colliding laser, $\ud \mcN_{\gamma}/\ud\omega_{\gamma}$ (dark, solid).
NBW becomes more probable as $\chi_{\gamma} \gtrsim 1$, and so in each case the value of the photon energy, $\omega_{\gamma}$, which satisfies $\text{max}[\chi_{\gamma}] = \eta_{\gamma} \xi_{\text{pairs}} = 1$ is shown (black, dashed).
For the lowest considered total laser energy, $E_{\text{total}} = 170$~J, one can see that only a small portion of the photons which collide with the secondary pulse are converted into electron-positron pairs.
The peak dimensionless intensity of the colliding laser in this case is $\xi_{\text{pairs}} \approx 111$, which means only photons with energies $\omega_{\gamma} \gtrsim 771$~MeV will experience peak values of the quantum nonlinearity parameter $\text{max}[\chi_{\gamma}] \gtrsim 1$. 
This corresponds to a only $\sim 10^{-5}\%$ of the total number of photons in the spectrum, or equivalently $\sim 10^{-4}\%$ of the total energy.
For the case $E_{\text{total}} = 510$~J, the peak dimensionless intensity increases to $\xi_{\text{pairs}} \approx 193$ and $\text{max}[\chi_{\gamma}] \gtrsim 1$ is satisfied for photons with $\omega_{\gamma} \gtrsim 445$~MeV, which accounts for $\sim0.2\%$ of the total number and $\sim 2.5\%$ of the total energy.
This then leads to a corresponding increase in the number of pairs produced.
The number of pairs then increases significantly for the highest energy case, $E_{\text{total}} = 1700$~J, where $\sim 10\%$ of the total number of photons ($\sim 35\%$ of the total spectrum energy) satisfy the condition $\omega_{\gamma} \gtrsim  244$~MeV which is required for $\text{max}[\chi_{\gamma}] \gtrsim 1$ with the dimensionless intensity parameter $\xi_{\text{flash}} \approx 352$.

\begin{figure*}[t!!]
    \centering
    \includegraphics[width=0.32\linewidth]{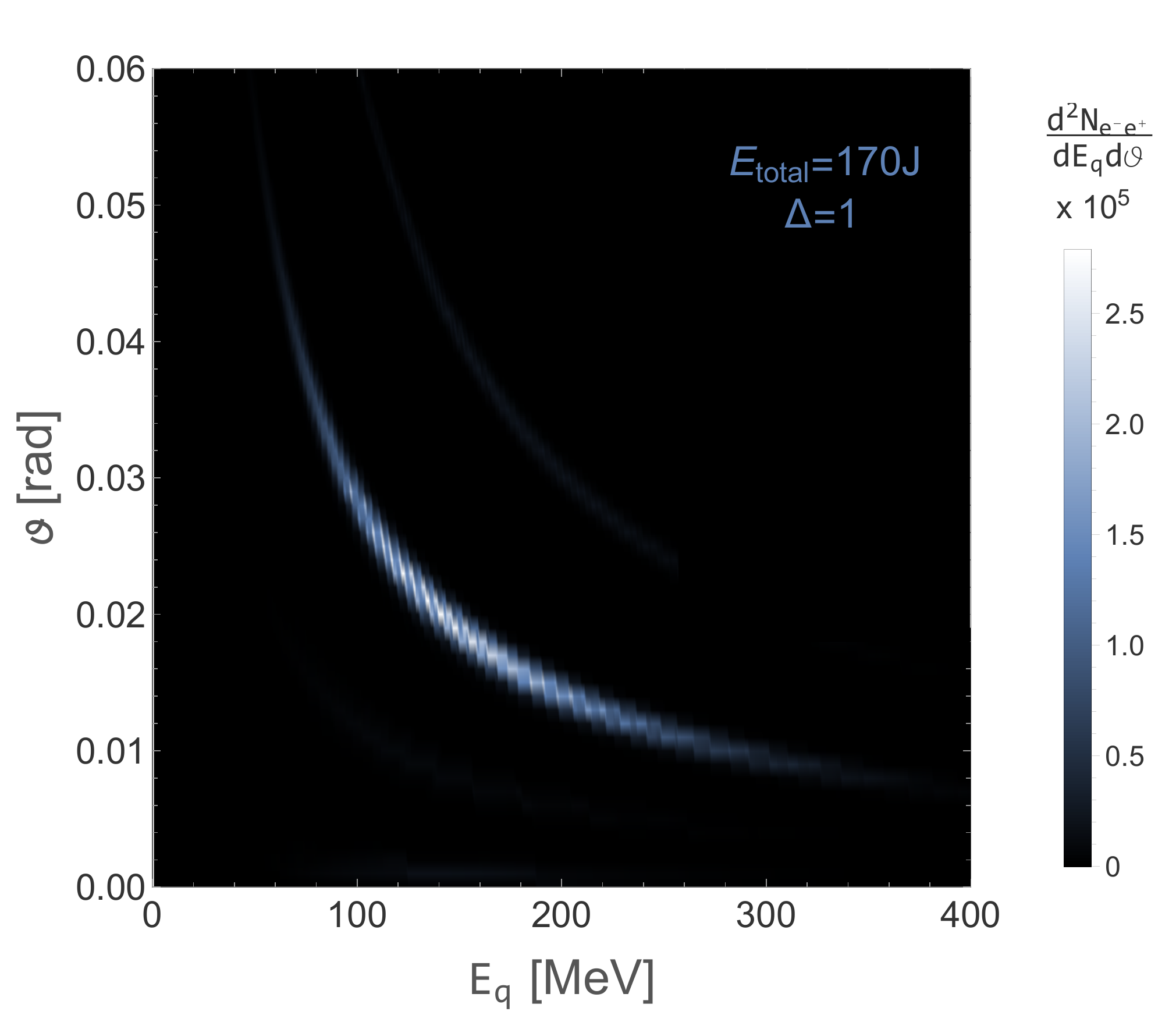}
    \includegraphics[width=0.32\linewidth]{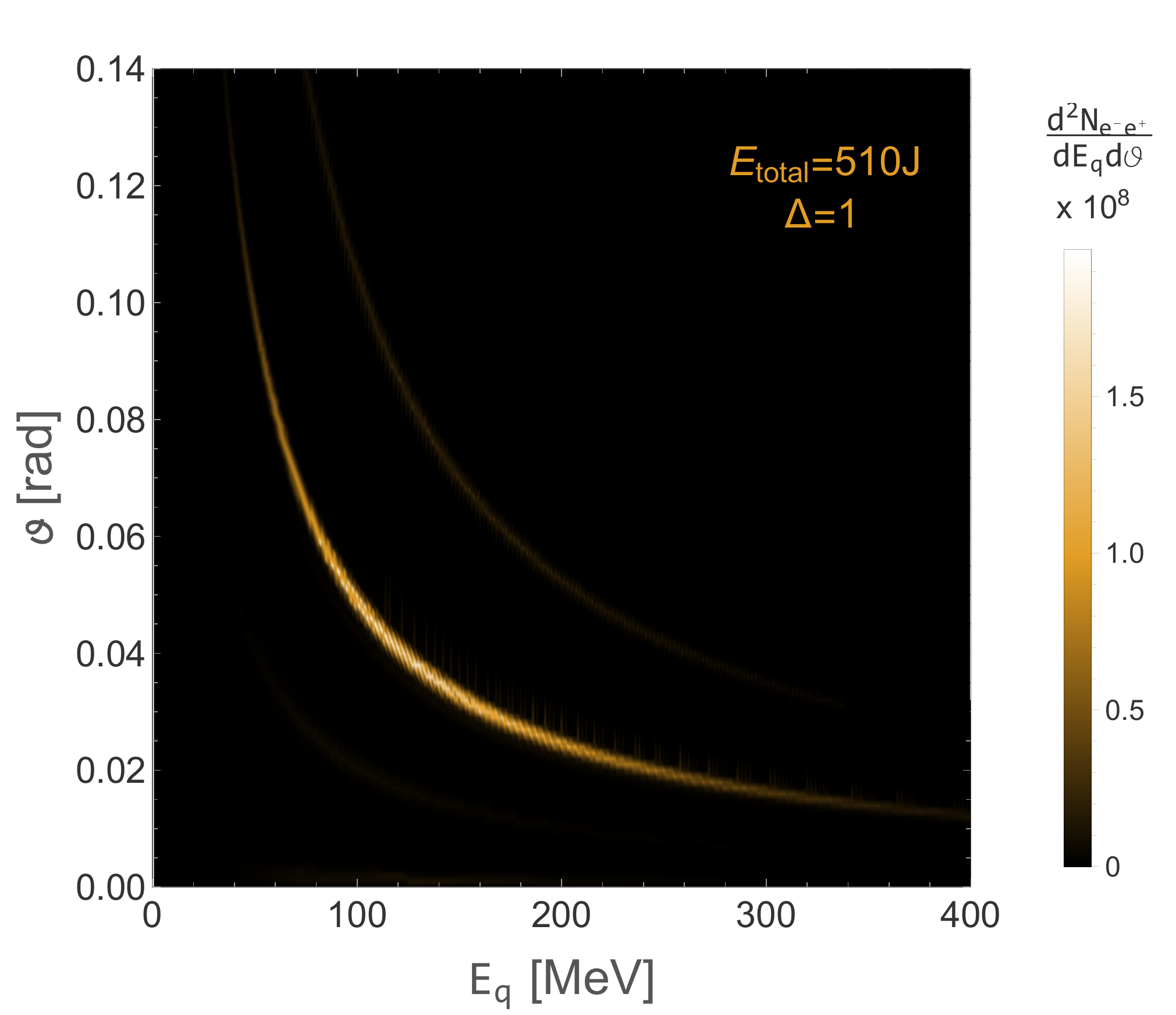}
    \includegraphics[width=0.32\linewidth]{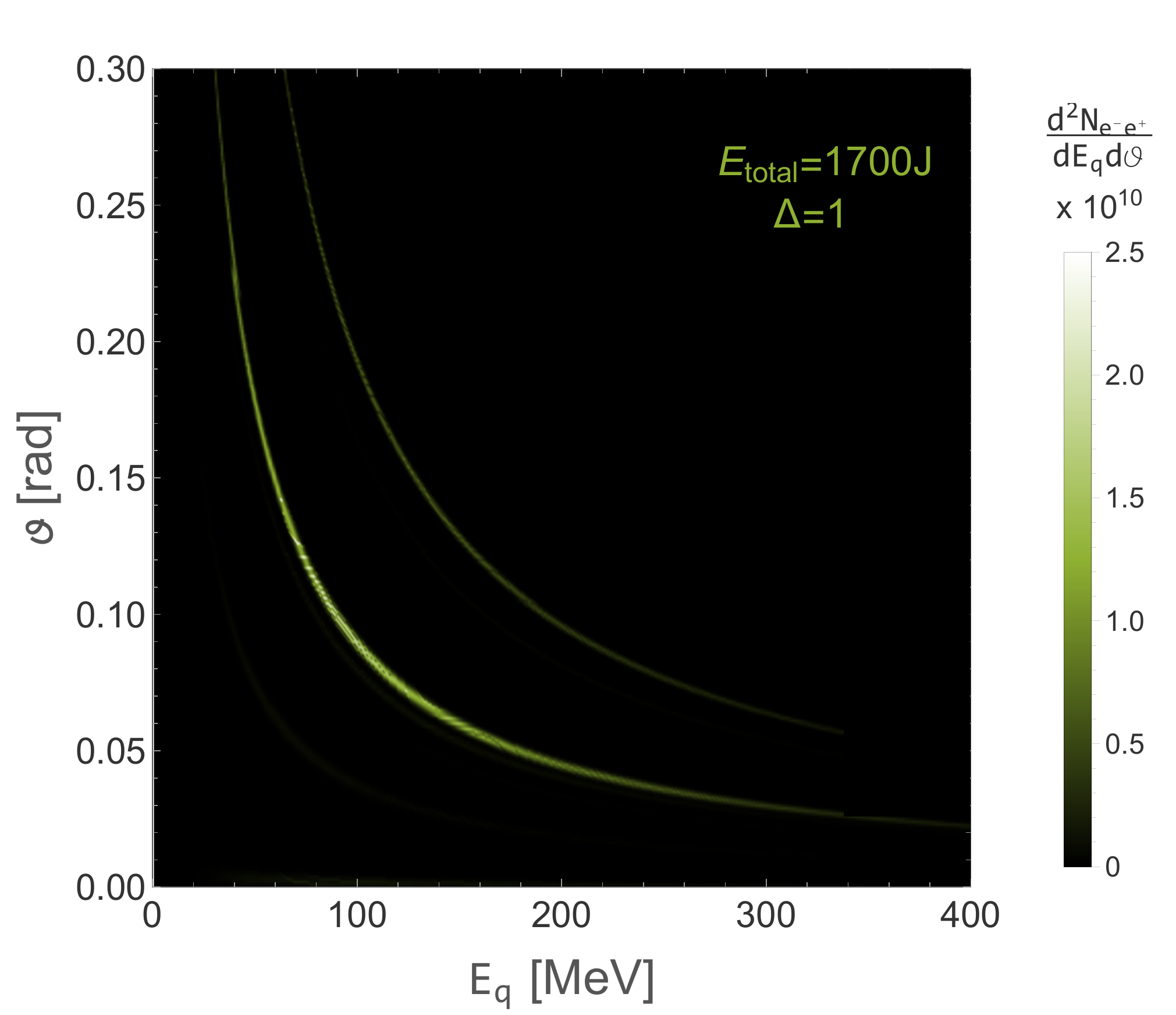}
    \caption{%
        Double differential spectra,
        $\ud^{2} \mcN_{e^{-}e^{+}}/\ud \vartheta \ud E_{q}$.
        \emph{Left:} $E_{\text{total}} = 170$~J. 
        \emph{Middle:} $E_{\text{total}} = 510$~J.                
        \emph{Right:} $E_{\text{total}} = 1700$~J.
    }
    \label{fig:Double}
\end{figure*}

\begin{figure*}[t!!]
    \centering
    \includegraphics[width=0.32\linewidth]{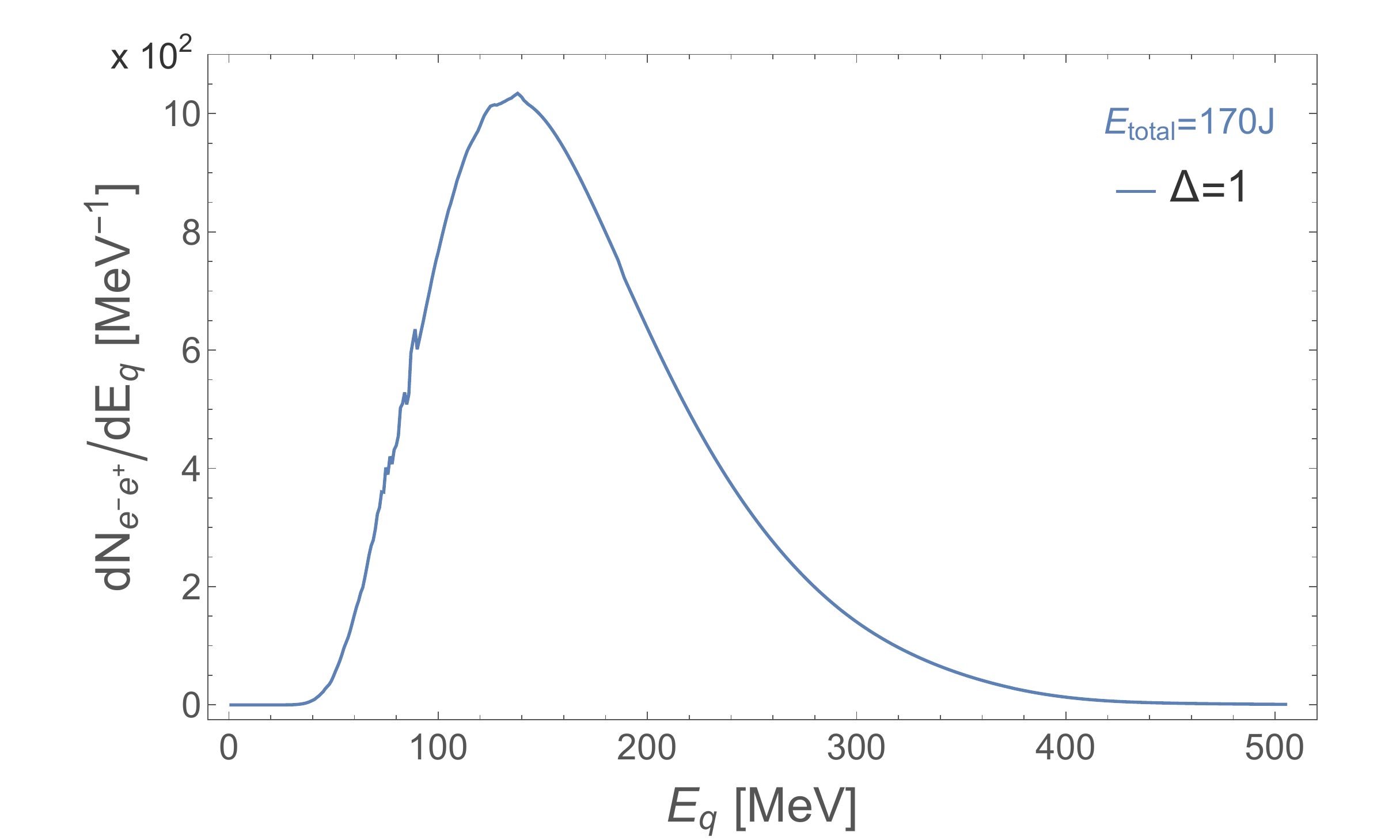}
    \includegraphics[width=0.32\linewidth]{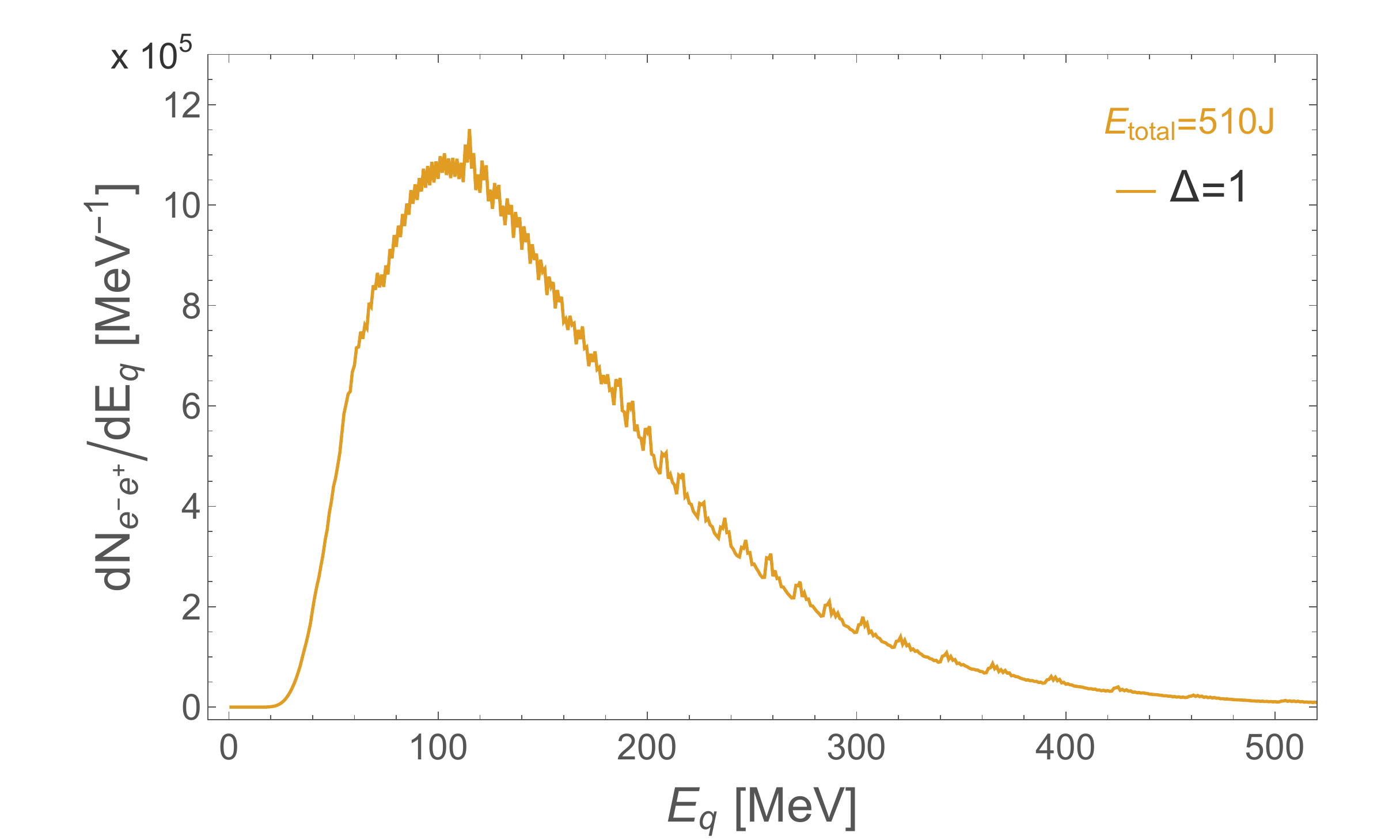}
    \includegraphics[width=0.32\linewidth]{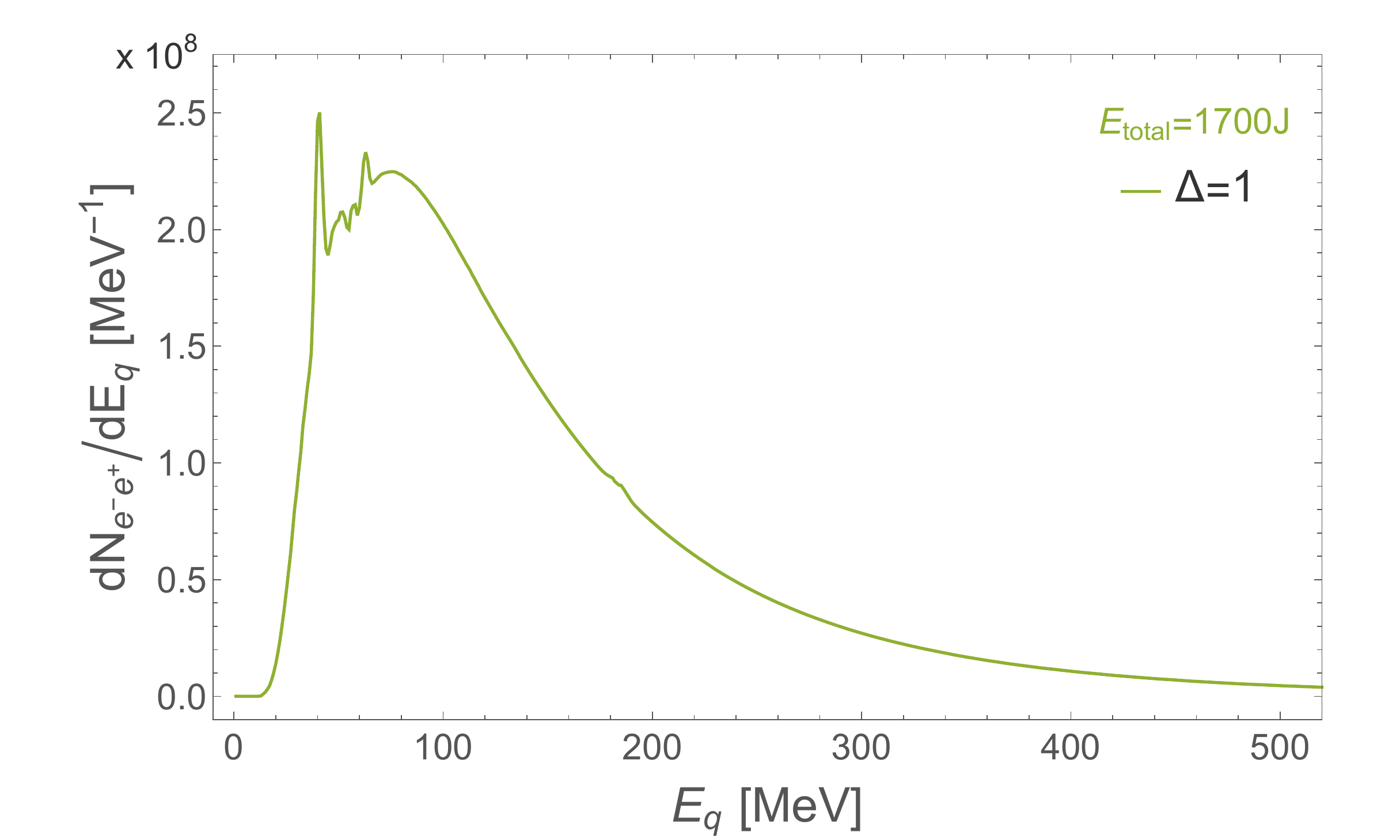}

    \includegraphics[width=0.32\linewidth]{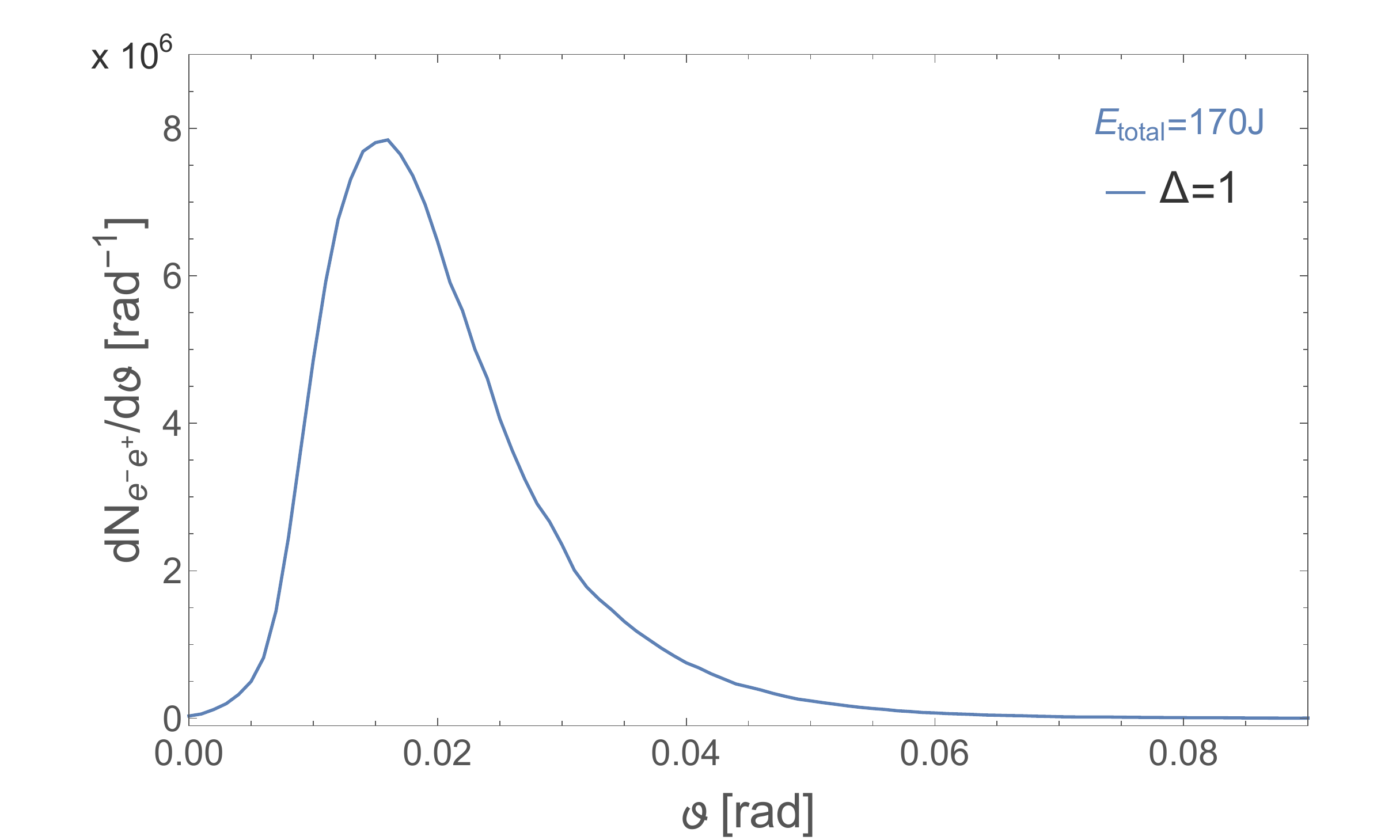}
    \includegraphics[width=0.32\linewidth]{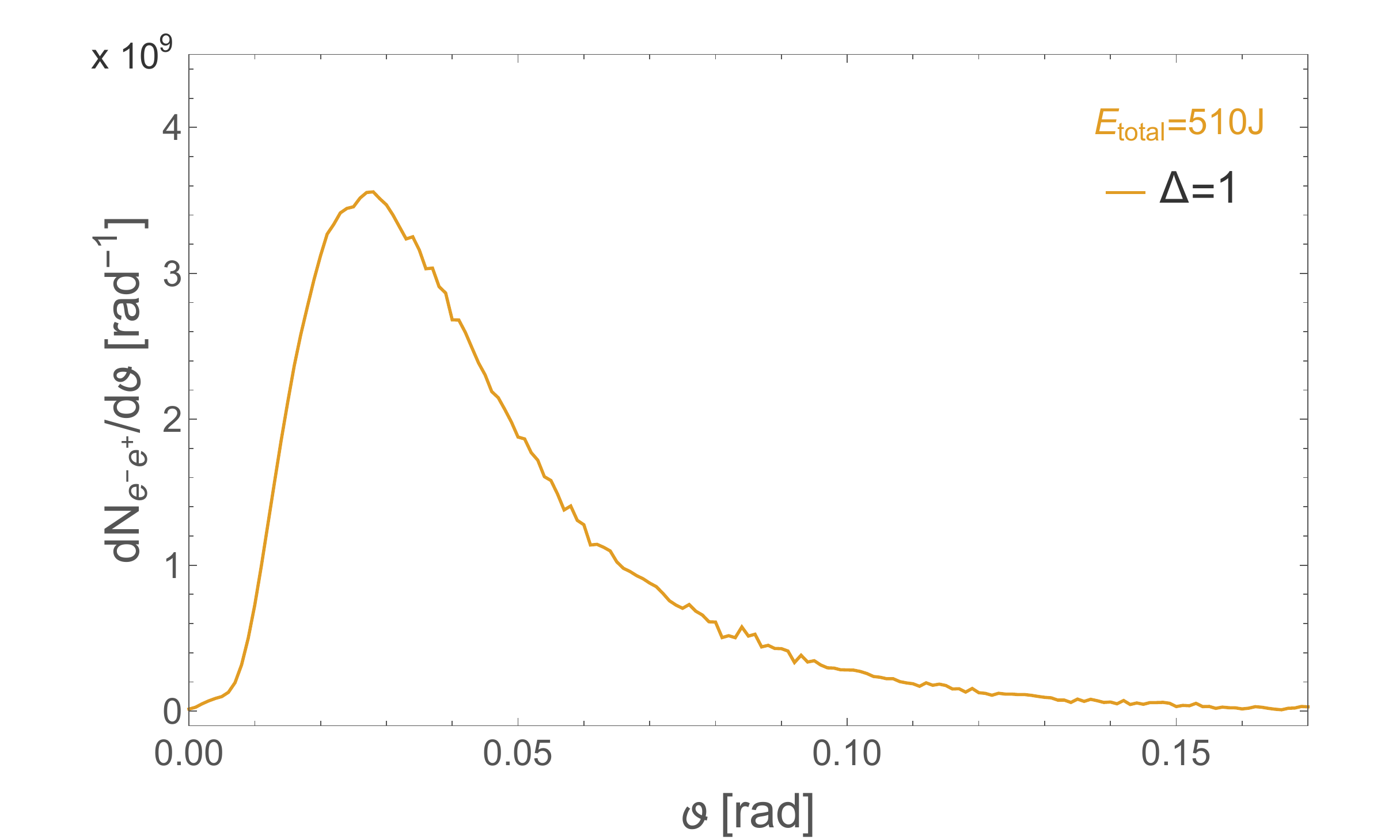}
    \includegraphics[width=0.32\linewidth]{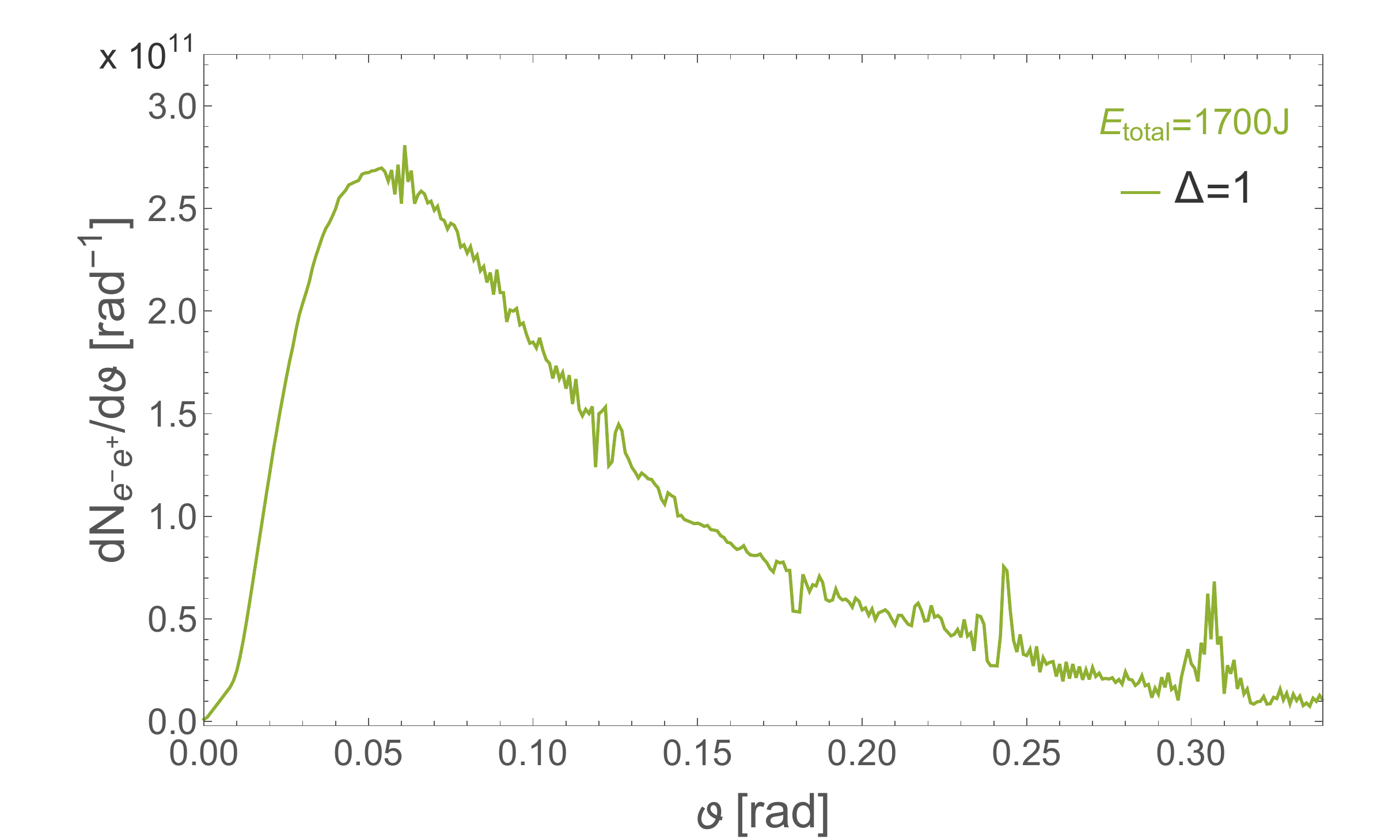}
    \caption{%
        Single differential spectra.
        Top row: energy spectra, $\ud \mcN_{e^{-}e^{+}}/\ud E_{q}$.
        Bottom row: angular spectra, $\ud \mcN_{e^{-}e^{+}}/\ud \vartheta$.
        \emph{Left:} $E_{\text{total}} = 170$~J, $\Delta=1$.
        \emph{Middle:} $E_{\text{total}} = 510$~J, $\Delta=1$.                
        \emph{Right:} $E_{\text{total}} = 1700$~J, $\Delta=1$.
    }
    \label{fig:Single}
\end{figure*}

Turning now to the properties of the produced positrons, \cref{fig:Double} shows the double differential spectra in the physical variables, $(\vartheta,E_{q})$, i.e. $\ud^{2} \mcN_{e^{-}e^{+}}/\ud\vartheta\ud E_{q}$, for the three cases of total laser energy, $E_{\text{total}}$, with $\Delta = 1$.
Similarly, \cref{fig:Single} shows the corresponding single differential energy and angular spectra, $\ud \mcN_{e^{-}e^{+}}/\ud E_{q}$ and $\ud \mcN_{e^{-}e^{+}}/\ud\vartheta$, respectively.
The structure of the NBW probability, \cref{eqn:AngularLCFA}, causes suppression of the pair production except around sharply peaked values of the dimensionless variable, $r$.
Increasing the total available laser energy leads to larger values of $r$ being attained by the produced electron-positron pairs, due to the stronger transverse field which they experience.
From $r \approx (E_{q}/m) \sin\vartheta$ we can see that increasing $r$ means increasing the energy of the produced positrons, $E_{q}$, and/or increasing their emission angle, $\vartheta$.
The double differential spectra in \cref{fig:Double} follow curved lines of approximately constant $r$, i.e. $\vartheta \approx \sin^{-1}(m r/E_{q})$.
Increasing the energy of the driving laser pulse increases both the maximum energy of the $\gamma$-flash photons and the overall number of photons, particularly those of lower energy (c.f.~\cref{fig:Spectra}).
As the energy of the secondary colliding pulse increases, more of the lower energy photons can be converted into electron-positron pairs due to the threshold energy for $\text{max}[\chi_{\gamma}] \gtrsim 1$ being reduced, see~\cref{fig:Spectra-Pairs-omega}.
This in turn leads to more low energy electron-positron pairs being produced, which pushes the peak in the energy spectrum to lower values of the energy, $E_{q}$, as the total available laser energy, $E_{\text{total}}$, is increased, as shown in the top row of \cref{fig:Single}.
Furthermore, low energy positrons are more strongly influenced by the electromagnetic field of the colliding laser, and are emitted at larger emission angles, $\vartheta$ (c.f. \cref{fig:Double}).
This leads to a broadening of the single differential angular spectra, $\ud\mcN_{e^{-}e^{+}}/\ud\vartheta$, as $E_{\text{total}}$ increases, as shown in the bottom row of  \cref{fig:Single}.

\section{Summary and outlook \label{sec:Summary}}

Current and next generation high-power laser facilities will be capable of delivering multi-PW peak power laser pulses for studying strong-field QED processes in the lab.
Here we have explored the feasibility of experimentally observing the nonlinear Breit-Wheeler process using a simple and efficient two-stage setup whereby high-energy photons produced by irradiating an overdense plasma with a laser pulse collide with a secondary laser pulse to produce electron-positron pairs.

Photons produced by the $\gamma$-flash mechanism have an angular spread, producing a double lobe pattern.
We found the optimal angle for the secondary collision by maximising the radiant intensity of the $\gamma$-flash photons, and showed that the photons within a full-angle divergence of $\beta = 10~\text{deg}$ around this optimal angle accounted for a large proportion of the highest energy $\gamma$-flash photons (c.f.~\cref{fig:Spectra}).

Considering three different cases for the available laser energy, $E_{\text{laser}} = (170,510,1700)$~J, which for pulses with duration $17$~fs corresponds to powers $P = (10,30,100)$~PW, we demonstrated that the optimal splitting  between the laser energy dedicated to producing photons via the $\gamma$-flash mechanism, $E_{\text{flash}}$, and the energy used in the secondary pulse responsible for converting photons to pairs via the nonlinear Breit-Wheeler mechanism, $E_{\text{pairs}}$, was $\Delta = E_{\text{flash}}/E_{\text{pairs}} = 1$.
In each case this optimised the number of electron positron pairs produced.
By locating the focus of the secondary laser pulse with energy $E_{\text{pairs}}$ at a distance $d = 10$~cm from the rear surface of the our findings suggest that with current capabilities on the order of 0.01 pairs could be produced per shot with $E_{\text{total}} = 170$~J, and next generation facilities capable of reaching 100~PW, with total available energy $E_{\text{total}} = 1700$~J could see as many as 1200 pairs being produced per shot.

As well as calculating the total number of produced pairs, we have considered the energy and angular spectra, showing that the broadness of the $\gamma$-flash energy spectrum produces positrons with a strong peak with high-energy tail.
For lower laser energies/powers the positrons are strongly emitted in the seed photon direction, but for larger energies/powers they can develop a larger transverse momentum component.
This is due to the higher laser energies allowing lower energy photons in the initial spectrum to decay into pairs, which are then more strongly kicked in the transverse plane by the laser.

Having demonstrated the feasibility of using the $\gamma$-flash mechanism to generate seed photons for nonlinear Breit-Wheeler pair production, there are a number of possible routes which could be explored in future work.
There are several optimisations which could be explored, both in the $\gamma$-flash photon production stage and the nonlinear Breit-Wheeler stage.
In the photon production stage we have kept the target thickness, pulse duration and focal spot size constant for each pulse energy, $E_{\text{flash}}$.
Related work which instead produces photons via bremsstrahlung from a high-energy electron colliding with a solid target~\cite{Blackburn.PPCF.2018} determined an optimal target thickness for maximising the number of electron-positron pairs produced in the second stage where the photons collide with an intense laser pulse.
In future work we will explore whether an optimal target thickness can also be determined in the $\gamma$-flash case.
This could lead to significant optimisations on two fronts.
Firstly, by increasing the energy and number of photons produced, which will directly increase the number of pairs which can be obtained in the second stage.
Secondly, increasing the target thickness can reduce the number of secondary charged particles produced by the $\gamma$-flash mechanism.
By minimising the number of background particles through an optimal thickness of the target and/or the duration of the laser pulse incident on the target, one could decrease the distance $d$ to the interaction point where the secondary laser pulse is focussed.
Since the total number of pairs produced scales with $d^{-2}$, this could lead to significant increases in the total number of pairs produced.
For example, for a distance $d = 1$~cm considered above, the estimated number of pairs per shot for the total laser energies $E_{\text{total}} = (170,510,1700)$~J are as high as $\mcN_{e^{-}e^{+}} \sim (1,1000,10^{5})$.

Another optimisation which could be explored further with regards to the photon generation stage is the use of alternative laser polarisations to reduce the divergence of the photon beam.
Here we have used a linearly polarised laser pulse, which produces the characteristic double-lobe angular intensity pattern in the $\gamma$-flash photons, but recent work~\cite{Hadjisolomou.JoPP.2021} utilising a radially polarised beam has indicated a much lower divergence of the photon beam.
This could lead to a significantly larger number of photons propagating to the focus of the secondary laser pulse, which could help increase the total number of pairs produced via the nonlinear Breit-Wheeler mechanism.

With regards to the second stage where the $\gamma$-flash photons collide with the second laser pulse of energy $E_{\text{pairs}}$, there are further improvements which will be explored in future work.
Firstly, we have chosen the beam waist of the secondary laser pulse to minimise the influence of focussing effects, and modelled the laser pulse as a plane wave with Gaussian temporal profile. 
This means that the highest intensities which the photons see in the laser pulse are not as high as could be achieved with stronger focussing.
However, it is known that focussing effects can have a detrimental effect on the number of pairs produced, see e.g~\cite{Blackburn.PPCF.2018,Mercuri-Baron.NJP.2021}.
Focussing effects can be included for high-energy photons and high field-strengths in the probabilities using, for example, a high-energy WKB approach~\cite{Piazza.PRL.2016}.
We will explore the interplay between the competing effects of reaching high field-strength and minimising the detrimental effects of focussing in future work.
Including focussing effects will also allow the effect of pulse duration to be considered beyond the plane wave model.

As noted previously, one of the parameters with the highest impact on the number of pairs produced is $d$, which is the distance from the source of the $\gamma$-flash photons to the focal spot of the secondary laser pulse.
Depending on the background of secondary charged particles produced in the $\gamma$-flash mechanism, this distance may need to be sufficiently large to allow for magnetic deflection of these particles using stationary magnets or other techniques (e.g.~\cite{Tilborg.PRL.2015}).
The pairs produced via the nonlinear Breit-Wheeler stage of the proposed set up have an angular spread.
In light of this it may be possible to to find a region in the angular plane where the background of charged particles is minimised, and the signal of nonlinear Breit-Wheeler pairs can be easily detected.
This would allow the distance, $d$, to be reduced without the need of magnetically deflecting the background particles.
This will require more accurate modelling of the final distribution of background particles at the detection region, for example by using additional simulations with QED-PIC codes, or Monte-Carlo codes such as FLUKA~\cite{Ahdida.Frontiers.2022}, Geant4~\cite{Allison.NIMPR.2016}, or similar.
We will consider this in future work.

\begin{acknowledgments}
    This work was supported by the project ``Advanced research using high intensity laser produced photons and particles'' (ADONIS) (CZ.02.1.01/0.0/0.0/16\_019/0000789) from the European Regional Development Fund.
\end{acknowledgments}

\bibliographystyle{apsrev}

\end{document}